\documentclass[aps,prl,twocolumn,citeautoscript,showpacs,notitlepage,superscriptaddress]{revtex4-1}

\usepackage[english]{babel}
\usepackage[pdftex,colorlinks=false,bookmarks=false]{hyperref}
\addto\extrasenglish{%

}%
\addto\captionsenglish{}
\usepackage{latexsym}
\usepackage{amsfonts}
\usepackage{amsmath,amssymb,amsthm}
\usepackage{bbm}
\usepackage{mathtools}
\usepackage[dvipsnames]{xcolor}
\usepackage{graphicx}
\usepackage{bm}
\usepackage{multirow}
\usepackage{tikz} 
\usepackage{txfonts}
\usepackage{ulem}
\usepackage{lipsum}
\usepackage{blindtext}
\usepackage{subfiles}
\usepackage{tikz-feynman}
\usepackage{amssymb}
\usepackage{amsmath}
\usepackage{eurosym}

\let\origunderline\underline
\renewcommand{\underline}[1]{\origunderline{\smash{#1}}}

\newcommand{\ul}[1]{\underline{#1}}
\newcommand{\bs}[1]{{\boldsymbol{#1}}}

\renewcommand{\d}{\mathrm{d}}
\renewcommand{\i}{\mathrm{i}}
\newcommand{\e}{\mathrm{e}}

\newcommand{\bk}{\bs{k}}
\newcommand{\bq}{\bs{q}}

\newcommand{\nF}{n_\mathrm{F}}

\renewcommand{\mathbb}{\mathbbm}

\newcommand{\pdd}[2]{\frac{\partial #1}{\partial #2}}

\newcommand{\eq}[1]{Eq.~\eqref{#1}}

\renewcommand{\dag}{\dagger}
\newcommand{\nodag}{{\vphantom\dag}}

\DeclareMathOperator{\tr}{tr}

\newcommand{\Qco}{Q_{\text{\scriptsize iCO}}}
\newcommand{\QCO}{\bs Q_{\text{\scriptsize iCO}}}
\newcommand{\qco}{\bs Q_{\text{\scriptsize CO}}}
\newcommand{\QCOt}{\tilde{\bs Q}_{\text{\scriptsize iCO}}}
\newcommand{\QSO}{\bs Q_{\text{\scriptsize SO}}}

\DeclareMathAlphabet\mathbfcal{OMS}{cmsy}{b}{n}

\renewcommand\d{\mathop{}\!\mathrm{d}}

\renewcommand{\i}{\mathrm{i}}

\newcommand{\psimf}{\scalebox{0.9}{$\langle\bm{\psi}\rangle $}}
\newcommand{\psimfs}{\scalebox{0.9}{$\langle\bm{\psi}\rangle $}}
\newcommand{\atmf}{\Bigg{|}_{\psimfs} }

\newcommand{\sz}{Z}

\newcommand{\szzb}{\underline{\bs \sz}}

\newcommand{\comma}{\ , }
\newcommand{\period}{\ . }

\newcommand{\bQ}{\bs{Q}}

\newcommand{\mueff}{\mu_\mathrm{eff}}
\newcommand{\muz}{\mu_0}

\renewcommand{\dag}{\dagger}

\newcommand{\ev}{\scalebox{1.025}{$\varepsilon$}} 
\newcommand{\evL}{\scalebox{1.3}{$\varepsilon$}}

\newcommand{\nfd}{n_{\scalebox{0.62}{F}}}

\newcommand{\QSOs}{\bs Q_{\text{\tiny SO}}}

\definecolor{copper}{rgb}{0.63, 0.29, 0.13} 
\definecolor{lightcopper}{rgb}{0.9704, 0.9432, 0.9304} 
\definecolor{dr}{rgb}{0, 0, 1}

\usepackage{cleveref}

\begin{document}
	
	\title{Interplay of spin and charge order in the electron-doped cuprates}
	
	\author{David Riegler}
	\email{david.riegler@physik.uni-wuerzburg.de}
	\affiliation{Institute for Theoretical Physics, University of Wuerzburg, D-97074 Wuerzburg, Germany}

	\author{Jannis Seufert}
	\affiliation{Institute for Theoretical Physics, University of Wuerzburg, D-97074 Wuerzburg, Germany}
	
	\author{Eduardo H. da Silva Neto}
	\affiliation{Department of Physics, Yale University, New Haven, CT 06511, USA}
	\affiliation{Energy Sciences Institute, Yale University, West Haven, CT 06516, USA}
	\affiliation{Department of Applied Physics, Yale University, New Haven, CT 06511, USA}
	\affiliation{Department of Physics, University of California, Davis, CA 95616, USA}
	
	\author{Peter W\"olfle}
	\affiliation{Institute for Theory of Condensed Matter and Institute for Quantum Materials and Technologies, Karlsruhe Institute of Technology, 76021 Karlsruhe, Germany}
	
	\author{Ronny Thomale}
	\affiliation{Institute for Theoretical Physics, University of Wuerzburg, D-97074 Wuerzburg, Germany}
	
	\author{Michael Klett}
	\email{michael.klett@physik.uni-wuerzburg.de}
	\affiliation{Institute for Theoretical Physics, University of Wuerzburg, D-97074 Wuerzburg, Germany}
	
	\date{\today }

	\begin{abstract}
		We study magnetic and charge order in the electron-doped high-$T_c$ cuprates based on the one-band Hubbard model with onsite ($U$) and nearest-neighbor $(V)$ interactions. To investigate the interplay between the orders, we employ the Kotliar-Ruckenstein slave-boson method and analyze fluctuations descending from an antiferromagnetic parent state.
		Our analysis reveals incommensurate charge order whose ordering vector matches the doping-dependence of resonant inelastic x-ray scattering (RIXS) measurements in Nd$_{2-x}$Ce$_x$CuO$_4$ (NCCO). 
		From our calculations of paramagnon dispersion as well as dynamical charge and spin structure factors, we reproduce all qualitative features of the RIXS signal.
	\end{abstract}

	\maketitle
	

	\textit{Introduction} ---In recent years, charge order (CO) has been found to complement spin order (SO) as an integral part of the low energy physics of cuprate superconductors \cite{tranquada1995evidence,RevModPhys.82.2421,ghiringhelli2012long,chang2012direct,tabis2014charge,comin2014charge,daSilvaNeto_2014,da2015charge,daSilvaNeto_2016,comin2016resonant,PhysRevB.98.161114,Kang_2019,arpaia2019dynamical,arpaia2021charge,wandel2022enhanced}. 
	Through various experiments showing either the suppression of CO with the onset of superconductivity or the enhancement of CO with a suppression of it, there is accumulating evidence on their competing character \cite{daSilvaNeto_2014,chang2012direct,tranquada1995evidence,wandel2022enhanced}.
	Still, dynamic fluctuations of CO-type pervade the entire cuprate phase diagram \cite{arpaia2019dynamical,arpaia2021charge,RevModPhys.87.457}, suggesting their quintessential role in cuprate phenomenology. The intertwining of CO and SO features important qualitative differences in electron- and hole-doped cuprates \cite{RevModPhys.82.2421}: The commensurate antiferromagnetic (AFM) phase extends to larger dopings in electron-doped cuprates whereas it becomes incommensurate in various hole-doped compounds, with significant impact on the nature of CO \cite{tranquada1995evidence,Yamada_1998,comin2016resonant}. 
	Furthermore, phase separation (PS) \cite{kivelson2003detect}, i.e., an inhomogeneous mixture of insulating magnetic and metallic domains, is well established for hole-doping but is assumed to be absent in electron-doped cuprates \cite{RevModPhys.82.2421}. 
	
	The one-band Hubbard model in two dimensions (2D) with on-site repulsion $U$ has been established as a minimal model to theoretically describe cuprate systems, and is assumed to also model finite hole- and electron-doping. Accounting for the dissimilarity of hole-doped and electron-doped cuprates, particle-hole asymmetry is already implied at the single-particle level by next-to-nearest neighbor (NNN) hopping $t'$ \cite{PhysRevB.71.134527,doi:10.1080/00018732.2014.940227,PhysRevB.98.134501,PhysRevB.72.054519}.
	\textit{Ab initio} calculations further imply a non-negligible nearest-neighbor (NN) repulsion $V$ \cite{PhysRevB.98.134501,PhysRevB.99.245155}, which is suggested to be relevant in capturing the phenomenology of static and dynamic charge order in hole-doped cuprates \cite{boschini2021dynamic,PhysRevB.106.224512,scott2023lowenergy}. In absence of an exact solution, a variety of different approximations and computational methods have been developed for the Hubbard model \cite{PhysRevX.5.041041}. Resolving the intertwined nature of SO and CO poses a particular numerical challenge, as it implies the necessity of addressing incommensurate order and dynamical correlation functions. 
	
	In this Letter, we develop a theory of intertwined charge and spin order for electron-doped cuprates. We adopt the Kotliar-Ruckenstein slave-boson (KRSB) ansatz \cite{kotliar_new_1986} which we have recently generalized to address a magnetic mean-field and dynamic fluctuations around the symmetry-broken state \cite{Seufert_2021}. We find that even moderate $V$ has a decisive impact on the electron-doped regime by removing its propensity to phase-separation. From a commensurate AFM parent state, we calculate the dynamical spin and charge susceptibility as a function of electron-doping. We discover a strong interdependence between charge fluctuations and the longitudinal spin channel. Enabled by KRSB to calculate the doping dependence of the ordering vectors, paramagnons, plasmons, and the dynamical spin and charge structure factors, we obtain good agreement with experimental findings in Nd$_{2-x}$Ce$_x$CuO$_4$ (NCCO) \cite{lee2014asymmetry,daSilvaNeto_2016,hepting2018three,PhysRevB.98.161114}. 
	
	
	\textit{Model and method} --- We employ the extended one-band $t$-$t'$ $U$-$V$ Hubbard model on the 2D square lattice with on-site repulsion $U$ and NN interaction $V$ defined by 
	
	\noindent
	\begingroup
	\allowdisplaybreaks
	\begin{align}\label{eq:Hamiltonian}
	\begin{split}
	H= &- \sum_{\sigma=\uparrow,\downarrow}
	\left(
	t^\nodag_{\vphantom i} \sum_{\langle i,j \rangle_1}  \, c_{i,\sigma }^{\dagger}c^\nodag_{j,\sigma}  
	+ t{\vphantom i}'  \sum_{\langle i,j \rangle_2} c_{i,\sigma }^{\dagger}c^\nodag_{j,\sigma}
	+\mathrm{h.c.}
	\right) \\
	&- \mu_0 \sum_i n_{i}  
	+U\sum_{i} c_{i,\uparrow }^\dagger c_{i,\uparrow}^\nodag c_{i,\downarrow }^\dagger c_{i,\downarrow }^\nodag 
	+ V \sum_{\langle ij\rangle_1} n_i n_j	\, , 
	\end{split}
	\end{align}
	\endgroup
	where the operator $c_{i,\sigma }^{\dagger }$ creates an electron with spin $\sigma =\{\uparrow ,\downarrow \} $ at site $i$, and $n_i= \sum_\sigma c_{i,\sigma }^{\dagger}c^\nodag_{i,\sigma} $. Moreover, $\langle i,j \rangle_n$ denotes a $n$th nearest neighbor pair, and $\mu_0$ is the chemical potential. 
	We employ $t'/t=-0.2$ throughout the paper, which is a generic choice to approximate a large family of cuprate materials \cite{doi:10.1080/00018732.2014.940227,PhysRevB.98.134501,PhysRevB.72.054519} and measure energy in units of $t$.
	
	\textit{Slave-boson mean-field approximation} ---We apply the spin rotation invariant Kotliar-Ruckenstein slave-boson (SRIKR-SB) representation ~\cite{kotliar_new_1986,woelfle_spin_1992}, whereby we introduce the bosonic fields $e_{i},d_{i},p_{0,i}$ and $\boldsymbol{p}_{i}$ which label empty, doubly and singly occupied states respectively, as well as a set of auxiliary fermionic fields $\bs f_{i}^{\nodag}$ and Lagrange multiplier fields $\alpha_i, \beta_{0,i}, \bs \beta_i$ to recover the physical subspace via constraints.
	We then determine the ground state within a static mean-field (MF) ansatz, where bosonic fields that couple to charge degrees of freedom, i.e., $\psi_{\mu,i} \in \{\alpha_i, e_i, d_i, p_{0,i}, \beta_{0,i}\}$ are approximated to be spatially uniform, whereas the spin degrees of freedom, i.e., $\psi_{\mu,i} \in \{\bs p_i, \bs \beta_i\}$ are constrained to a spin spiral with ordering vector $\QSO$. The resulting MF expectation values of the charge and spin operator are given by
	$n_i\rightarrow \langle n \rangle =1 +d^2-e^2$ and $\bs S_i\rightarrow \langle \bs S_i \rangle =p_0 |\bs p|\left[ \cos(\QSO  \bs r_i), \sin(\QSO \bs r_i),0 \right]$. The MF ground state is thereby determined by the saddle point of the Free Energy, where the constraints are enforced on average.
	As we derive in the supplemental material (SM) \cite{SM}, the addition of $V$ shifts the chemical potential $\mu_0$ and Lagrange-multiplier $\beta_0$
	\begin{equation}\label{eq:muz:p}
	\mu_0 =   \mu_0 \Big{|}_{V=0} + 4V n  , \quad 
	\beta_0 =   \beta_0\Big{|}_{V=0} + 4V n \, ,
	\end{equation}
	whereas all other MF variables are agnostic to $V$. The MF band structure remains unchanged since it only depends on $\mu_0-\beta_0$ \cite{Hubbard_Wuerzburg,Riegler2022,Klettthesis}.
	
	\textit{Gaussian fluctuations} ---
	Beyond mean-field, we consider fluctuations around the saddle point through expanding the action $\mathcal{S}$ up to second order in bosonic fields, i.e., calculating
	\begin{equation}
	\mathcal{M}_{\mu\nu}^{ab}(\bs q,\i\omega_n)
	=\frac12 \frac{\delta^2 \mathcal{S}(\psi)}{\delta \psi_{\mu,-\bs q - a\QSO, -\i\omega_n} \, \delta \psi_{\nu,\bs q+b\QSO, \i\omega_n}} \, .
	\end{equation}
	The established formalism \cite{Li1991,woelfle_spin_1997} has recently been extended to encompass symmetry-broken states \cite{Seufert_2021},
	which we apply to calculate susceptibilities descending from AFM mean-field saddle points, i.e., $\QSO=(\pi,\pi)$ with the Umklapp momenta $a,b\in \{0,1\}$. 
	The spin susceptibility is composed of the longitudinal part parallel to the staggered MF magnetization $\chi^l_s(\bs q, \omega) = \langle \delta S^x_{\bs q,\omega}\,  \delta S^x_{-\bs q,-\omega} \rangle$
	that couples to the charge susceptibility $\chi_c(\bs q, \omega) = \langle \delta n_{\bs q,\omega} \, \delta n_{-\bs q,-\omega} \rangle$ and a perpendicular, i.e., transversal part $\chi^t_s(\bs q, \omega) = \langle \delta S^y_{\bs q,\omega}\,  \delta S^y_{-\bs q,-\omega} \rangle=\langle \delta S^z_{\bs q,\omega}\,  \delta S^z_{-\bs q,-\omega} \rangle$ that is decoupled due to a block-diagonality in $\mathcal{M}^{ab}_{\mu\nu}$. Notably, we show in the SM along with a recap of the SB theory that the newly implemented interaction $V$ generates additional contributions for the charge fields $(e,d_1,d_2)$ in the bosonic part of the fluctuation matrix \cite{SM}.
	
	
	\textit{Mean-field analysis} ---
	In agreement with previous SB studies of the Hubbard model \cite{Fresard_1992,igoshev2015,Hubbard_Wuerzburg,Riegler2022,Klettthesis}, we determine incommensurate spiral magnetism for hole-doping
	$\big[\QSO/2\pi=\big(\frac12 ,\frac12-\delta\big)$ along with $\QSO/2\pi= \big(\frac12-\delta,\frac12-\delta\big) \big]$
	and an extended commensurate AFM domain for electron-doping $\big[\QSO/2\pi=\big(\frac12,\frac12\big)\big]$, which is further supported by other theoretical methods \cite{PhysRevB.93.035126,HMStripe}. 
	Moreover, the incommensurability $\delta$ is proportional to the doping $x$ \cite{igoshev2015,Hubbard_Wuerzburg}, and in fact in agreement with the Yamada-relation $\delta \approx x$ \cite{Klett2023} known from La-based cuprate families \cite{Yamada_1998,PhysRevB.78.212506,lee2022generic}.
	The doped magnetic MF ground state, however, is known to become unstable beyond a critical interaction \cite{PhysRevB.81.094407,igoshev2015,Seufert_2021,crispino2023slave}, which is signaled by a negative compressibility $\kappa_T \propto \partial n/\partial \mu_0 < 0$. Thereby, $\kappa_T^{-1} =\chi_c^{-1}(\bs q=0) = 0$ describes the tipping point where the MF ground state is no longer stable against charge fluctuations. 
	We find that this is avoided in \eq{eq:Hamiltonian} by the formation of two coexisting, individually stable condensates with different fillings and the same chemical potential, i.e., phase separation (PS):	
	The undoped, strongly magnetized and insulating AFM mixes with a metallic state carrying the doped holes or electrons with a lower net magnetization.
	While a weakly-interacting system tends to delocalize its charge carriers uniformly to gain kinetic energy, the potential energy scale set by the strong effective AFM Heisenberg exchange of the half-filled Hubbard model becomes dominant at low dopings. 
	This is why the PS scenario is favored over a uniform AFM, as it partially recovers the pristine crystal \cite{Seufert_2021,Klettthesis,Riegler2022}.
	For hole-doping, we associate PS obtained from our MF theory with the much discussed stripe phases \cite{kivelson2003detect,comin2016resonant} and find the ground state energies \cite{Riegler2022} to be comparable to those of an SB cluster study with large unit cells \cite{Raczkowski_2006}.
	In \eq{eq:Hamiltonian}, repulsive $V$ drastically reduces phase separated domains, which is due to the chemical potential shift described by \eq{eq:muz:p} and can intuitively be understood by the thereby effectively reduced AFM Heisenberg exchange. In particular, PS is found to be comparably resilient for hole-doping, whereas already rather small values of $V$ eliminate this ordering tendency for electron-doping as further elaborated in the SM \cite{SM}.
	In agreement with this emergent particle-hole asymmetry in our theory, PS has so far only been observed in hole-doped cuprate families \cite{RevModPhys.82.2421,RevModPhys.75.473,condmat4040087,PhysRevB.101.125107,PhysRevB.64.220507}. Moreover, for NCCO, precise measurements of core-level photo-emission spectra have shown that $\mu_0(n)$ is a monotonous function, i.e., that PS should be absent \cite{PhysRevB.64.220507}.

\begin{figure}[t!]
		\includegraphics[width=0.495\textwidth]{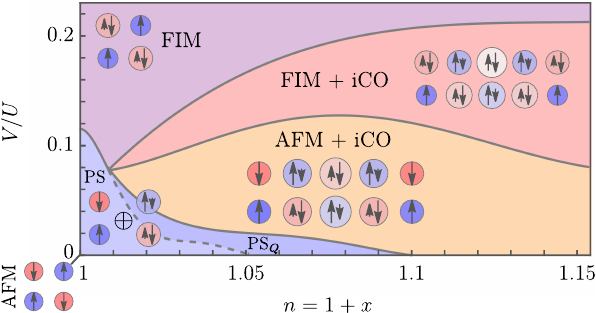}
		\caption
		{
			Phase diagram of emerging CO inferred from fluctuations around AFM mean-field ground states as a function of the interaction ratio $V/U$ and electron-doping $x$ in the low temperature limit. Different orders are visualized by idealized sketches, depicting how doped charges arrange in a real-space picture that combines magnetic ($\uparrow,\downarrow$) and charge (circle size) order.
		}
		\label{fig:PD}
	\end{figure}

	\textit{Fluctuation analysis} --- Due to this fragility of PS for electron-doping, we are particularly interested in other types of emerging charge inhomogeneities in the associated commensurate AFM domain and apply the fluctuation formalism for a more in-depth analysis. Hence, we track divergences of the momentum-dependent static charge susceptibility, i.e., $1/\chi_c(\bs q= \qco,\omega=0)=0$ that implicate a second order phase transition as function of the electron-doping $x$ and interaction ratio $V/U$. We define the leading instability by the divergence of $\chi_c$ that occurs at the lowest interaction value $U=U_c$ and classify the type of emerging long-range CO by the respective ordering vector $\qco$. The resulting phase diagram is shown in \autoref{fig:PD} and features four distinct types of CO.
	
	Small interaction ratios $V/U$ and dopings $x$ promote PS in full consistency with the previously discussed MF picture [i.e., $\qco=(0,0)$]. At intermediate dopings, the ordering vector evolves to small but finite values [$\qco=(\epsilon,0), \, \epsilon/2\pi \lesssim 0.125$] within a second order transition, whereby the precise value of $\epsilon$ depends on $V$. This type of order, labeled $\mathrm{PS}_{\boldsymbol Q}$, has been discussed in Refs. \cite{Seufert_2021,Riegler2022,Klettthesis} and interpreted as emerging domain walls, driven by the same physical mechanism as PS.
	Large ratios $V/U$ induce $\qco=(\pi,\pi)$, implicating a checkerboard charge density modulation on top of the AFM spin order. This apparent spin-charge coupling gives rise to an accompanying divergence of $\chi_s^l$ at the $\Gamma$-point, i.e., the onset of ferrimagnetism (FIM) with two different anti-parallel sublattice magnetizations. The emergence of such checkerboard CO and SO is further supported by previous slave-boson MF studies of the extended Hubbard model \cite{deeg1993slave,philoxene2022spin}. Notably, we find FIM and AFM ordering tendencies to compete at intermediate ratios $V/U\approx 0.1$, which results in an overall enhanced critical interaction scale $U_c$ as further elaborated in the SM \cite{SM}.
	
\begin{figure}[t!]
		\includegraphics[width=0.49\textwidth]{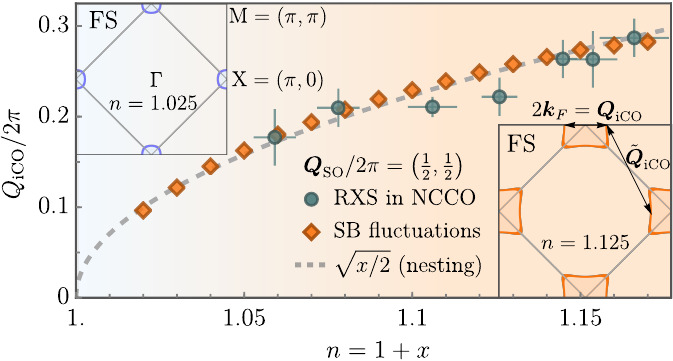}
		\caption{
			Incommensurate charge order (iCO) wave vector inferred from SB fluctuations in comparison to RXS data in NCCO adopted from Ref.~\cite{daSilvaNeto_2016}. Our result for $\QCO$ follows the diameter of the Fermi surface (FS) that relates to the doping $x$ via Luttinger's theorem (dashed line).			
		}
		\label{fig:QCO}
	\end{figure}

	With increasing doping the kinetic energy gains importance, whereby we detect emerging incommensurate charge order (iCO) with ordering vectors $\bs \QCO=(\Qco,0)$ as modulation of the AFM and $\bs \QCOt=(\pi-\Qco,\pi)$ of the FIM, respectively. The resulting orders are sketched in \autoref{fig:PD}, albeit somewhat idealized and truncated to a finite unit cell, while in general the pattern remains incommensurate in one spatial direction. We identify both modulations as a result of Fermi surface instabilities, i.e., kinetic effects due to nesting (intra-pocket vs.~inter-pocket) within the fermiology of the AFM parent state in the extended zone scheme as illustrated in the inset of \autoref{fig:QCO}. 
	
	\textit{Charge order in NCCO} ---
	Emerging CO that partially overlaps with AFM order has been investigated via resonant x-ray scattering (RXS) experiments by measuring $\QCO$ in various electron-doped cuprates \cite{da2015charge,daSilvaNeto_2016,PhysRevB.98.161114,Kang_2019,hepting2018three,comin2016resonant}. Among those materials, NCCO also employs an extended AFM domain up to dopings of $x\lesssim 0.14$ \cite{motoyama2007spin} in agreement with the SB mean-field ground state. 
	Moreover, the FS is presumed to play an important role in determining the CO wave vector and its doping dependence \cite{da2015charge,daSilvaNeto_2016,Kang_2019} but other mechanisms like a momentum-dependent electron-phonon coupling, which is not captured by our model, could also increase the propensity towards CO \cite{johannes2008fermi,deeg1993slave}.
	\autoref{fig:QCO} shows $\Qco$ inferred from the SB fluctuation calculation and RXS measurements in NCCO with rather good agreement. 
	Within our theory, we can clearly establish the association $\Qco=2k_F$ between the CO wave vector and the FS, which implies the CO to be nesting-enhanced.
	According to Luttinger's theorem the enclosed area of the FS is equal to the doping $x$, whereby we estimate $2k_{\text{F}}/2\pi \approx \sqrt{x/2}$ under the approximation of square-shaped FSs, which matches the	CO wave vector.
	
	The emerging CO already becomes apparent below the critical interaction $U_c$ in the form of short-range correlations, implied by a finite peak of $\chi_c$ at the same wave vector $\QCO$ that is mostly independent of the interactions $U,V$ \cite{SM}. Indeed, the correlation length extracted from RXS measurements in NCCO is estimated to be rather short-ranged with $\xi\approx 5$ unit cells \cite{daSilvaNeto_2016,Kang_2019} and it has been shown that the CO correlations at $\QCO$ remain visible beyond the AFM Néel temperature $T_{\text{\scriptsize Néel}}$ for sufficiently high dopings $x\gtrsim 0.08$ \cite{daSilvaNeto_2016}. In line with this, these are also present within the paramagnetic (PM) domain of our theory, where $\chi_c$ employs a finite nesting-peak at $\QCO$. Back-folding of the FS at the transition from PM to AFM order enhances this type of nesting for electron-doping. 	
	For hole-doping, the situation is fundamentally different. While there is a nesting-peak on the $\Gamma$-$\mathrm{X}$ high symmetry line similar to electron-doping within PM ground states, that type of nesting is completely removed within the AFM as shown in the SM \cite{SM}. 
	
	Finally, we discuss dynamic excitations and the integrated structure factors
	\begin{equation}\label{eq:corr}
	S_{c,s}(\boldsymbol q) =1/\pi \int_{\Omega_\text{min}}^{\Omega_\text{max}} d\omega \ \text{Im}\,\chi_{c,s}(\boldsymbol q,\omega)   \left[ 1+n_{\text{B}}(\omega) \right] 
	\end{equation}
	with $n_{\text{B}}(\omega)=1/(\e^{\omega/T}-1)$, which can directly be related to resonant inelastic x-ray scattering (RIXS) \cite{RevModPhys.83.705}, which is energy-resolved in comparison to RXS. Thereby, da Silva Neto \textit{et.al} revealed that roughly half of the signal at $\QCO$ can be attributed to the quasi-elastic line, i.e., $\omega=0$, which supports the claim of static CO. Additional high energy dynamic contributions near $\QCO$, appear mainly in the crossed-polarized channel, i.e., are indicated to be magnetic in nature and follow the paramagnon dispersion \cite{PhysRevB.98.161114}.

\begin{figure}[t!]
		\includegraphics[width=0.485\textwidth]{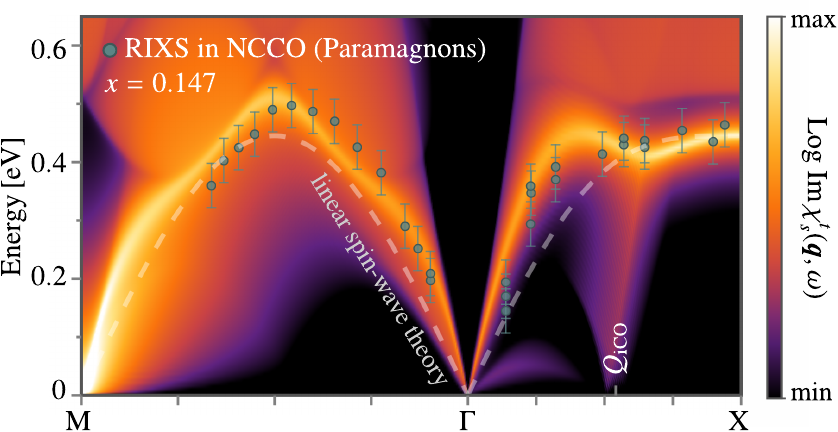}
		\caption{Paramagnon dispersion on the high-symmetry-path, which appears as enhanced line of $\chi_s^t$ ($U=5.5, V/U=0.1$), along with RIXS data in NCCO adopted from Ref.~\cite{lee2014asymmetry}. 
			The absolute energy scale in our model is fitted to $t=0.52$~eV. The dashed line displays linear spin-wave theory with the velocity $c_s=1.24~\mathrm{eV}\AA$ adopted from Ref.~\cite{lee2014asymmetry}.}
		\label{fig:paramagnons}
	\end{figure}

	
	Within our theory, we detect (para-)magnons in the transversal spin channel $\chi_s^t$. At low dopings, we find their dispersion to be sharp and to match linear spin-wave theory \cite{SM}, which is the signature of magnons and in quantitative agreement with RIXS measurements in NCCO \cite{lee2014asymmetry}. 
	With increasing doping we observe a broadening of the mode, which reflects the reduced correlation length and lifetime of paramagnons. 
	Moreover, as shown in \autoref{fig:paramagnons} for $x=0.147$, we find the paramagnon dispersion 
	in agreement with experimental evidence \cite{lee2014asymmetry} to deviate from linear spin-wave theory, which is displayed by the dashed line.
	Notably, we determine a nesting-enhanced, non-collective mode around $\QCO$ that merges with the paramagnon branch at $E\approx 400$~meV, where the experimentally measured high energy contributions to $\QCO$ are centered for that doping. The respective RIXS data \cite{PhysRevB.98.161114} suggests that this effect may also be present in NCCO but polarimetric measurements with higher energy resolution are needed for confirmation.
	
	Besides, RIXS experiments revealed a gapped collective mode around the $\Gamma$-point \cite{lee2014asymmetry} that has been identified as charge-mode and is also present in  La$_{2-x}$Ce$_x$CuO$_4$ (LCCO) \cite{hepting2018three}. In qualitative agreement, we detect such a mode in the charge channel but the gap size does not match quantitatively, compare SM \cite{SM}. 
	This discrepancy could be due to a three-dimensional nature of the charge mode. Measurements revealed the gap size to depend on the out-of-plane momentum $q_z$, whereas the paramagnons did not disperse appreciably along $q_z$ \cite{hepting2018three}.

\begin{figure}[t!]
		\includegraphics[width=0.49\textwidth]{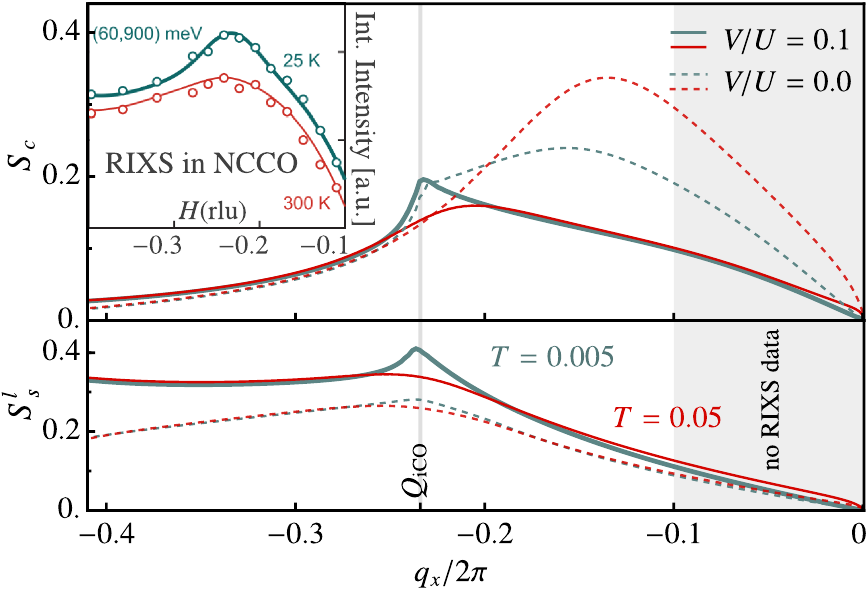}
		\caption
		{
			Charge ($S_c$) and longitudinal spin ($S_s^l$) structure factor for $x=0.108$ and $U \approx U_c$, i.e., $U=8.7$ for $V=0$ and $U=11.5$ for $V/U=0.1$ as function of $\bq=(q_x,0)$ for the energy cutoffs $\Omega_{\text{min}}=0.01$, $\Omega_{\text{max}}=2$. The inset shows RIXS data in NCCO adopted from Ref.~\cite{PhysRevB.98.161114} with the same doping and comparable energy cutoffs.
		}
		\label{fig:Sc}
	\end{figure}

	
	In contrast to $\chi_s^t$, the charge ($\chi_c$) and longitudinal ($\chi_s^l$) spin channel feature significant spectral weight at $\QCO$ for $\omega=0$. However, a direct measurement of the elastic line is subtle because soft x-ray RIXS is done in a reflection scattering geometry, which leads to an increased signal for small momenta. Therefore, we compare our results for the respective structure factors at doping $x=0.108$ in \autoref{fig:Sc} with RIXS data that includes small energies but not strictly the elastic line.
	The signal shown in the inset contains contributions from the charge and spin channel with integrated energies $E \in (60,900)$~meV and $\Delta E \approx 60$~meV \cite{PhysRevB.98.161114}. If we gauge our energy scale with \textit{ab initio} calculations in NCCO, i.e., $t\approx0.42$~eV \cite{PhysRevB.72.054519}, the shown  temperatures in the theoretical and experimental data are comparable. 
	Notably, Ref.~\cite{PhysRevB.98.161114} also revealed a coupling between dynamic magnetic and charge correlations in line with our theoretical predictions.
	As a result, $S_c$ and $S_s^l$ feature a coinciding peak at $\QCO$ that significantly flattens at room temperature like the RIXS signal. This qualitative agreement, however, requires a finite NN interaction $V$ that reduces the spectral weight in the charge channel at low momenta and enhances the signal at $\QCO$ in the longitudinal spin channel. For $V=0$, we predict a second excitation peak in $S_c$ due to the propensity towards phase separation ($\mathrm{PS}_{\bs Q}$), which is not supported by the available experimental data. 
	According to \textit{ab initio} calculations in the cuprates \cite{PhysRevB.106.235150} even longer-ranged interactions
	could be non-negligible and extend the experimentally observed AFM+iCO state by further suppressing PS and weakening FIM ordering tendencies.
	
	\textit{Conclusion} --- We identify the electron-doped Hubbard model to feature nesting-enhanced incommensurate charge order on top of a commensurate AFM background. Equipped with dynamical correlation profiles received from our slave-boson fluctuation analysis, we achieve quantitative correspondence with experimental evidence from resonant inelastic x-ray scattering. Our study resolves the intricacy of intertwined spin and charge order in cuprates, and highlights the use of slave-boson theories for multi-order phenomena in correlated electron systems.

\begin{acknowledgments}
		\textit{Acknowledgments}--- 
		The authors thank Werner Hanke, Masa Imada and Giorgio Sangiovanni for helpful discussions. The work in W\"urzburg is funded by Deutsche Forschungsgemeinschaft (DFG, German Research Foundation), Project-ID 258499086 - SFB 1170 and the W\"urzburg-Dresden Cluster of Excellence on Complexity and Topology in Quantum Matter -- \textit{ct.qmat} Project-ID 390858490 - EXC 2147. Peter W\"olfle acknowledges support through a Distinguished Senior Fellowship of Karlsruhe Institute of Technology. 
		Eduardo H. da Silva Neto acknowledges support by the Alfred P. Sloan Fellowship and the National Science Foundation, Grant No. 2034345.
	\end{acknowledgments}

	\bibliography{bibliography}
	
\newpage
\clearpage

\onecolumngrid

\begin{center}
	\textbf{\large Supplemental Material: Interplay of spin and charge order in the electron-doped cuprates}
\end{center}

\setcounter{equation}{0}
\makeatletter
\renewcommand{\theequation}{S\arabic{equation}}

\setcounter{section}{0}
\makeatletter
\renewcommand{\thesection}{\Roman{section}.}
\let \ds@sec=\section
\renewcommand{\section}[1]{\refstepcounter{section}\ds@sec{\thesection\ #1}}

\setcounter{subsection}{0}
\renewcommand{\thesubsection}{\Alph{subsection}.}
\let \ds@subsec=\subsection
\renewcommand{\subsection}[1]{\refstepcounter{subsection}\ds@subsec{\thesubsection\ #1}}

\setcounter{subsubsection}{0}
\renewcommand{\thesubsubsection}{\arabic{subsubsection}.}
\let \ds@subsubsec=\subsubsection
\renewcommand{\subsubsection}[1]{\refstepcounter{subsubsection}\ds@subsubsec{\thesubsubsection\ #1}}

In \cref{sec:SBMF}, we provide key results of the established Kotliar-Ruckenstein slave-boson (SB) mean-field (MF) theory \cite{kotliar_new_1986,woelfle_spin_1992,Fresard_1992,woelfle_spin_1997,Hubbard_Wuerzburg,Seufert_2021,Klettthesis,Riegler2022} and derive a consistent MF treatment of the nearest-neighbor (NN) interaction $V$, which has been introduced in the main text.
\Cref{sec:SBfluc} recaps the established SB fluctuation formalism \cite{woelfle_spin_1997,Hubbard_Wuerzburg} that has recently been extended to encompass magnetic saddle points \cite{Seufert_2021}
and derives additional contributions to the fluctuation matrix that arise due to the newly-implemented interaction $V$, where the final result given by \eq{eq:MV:result}.  In  both \cref{sec:SBMF,sec:SBfluc}, we adopt the compact notation of \cite{Riegler2022}, which also includes a more detailed review of the method. 
Subsequently, \cref{sec:maxwell} exemplifies Maxwell constructions that are performed to detect MF phase separation (PS) and \cref{SM:sec:fermiology} discusses differences in MF fermiology between electron- and hole-doping. In \cref{sb:sec:uc}, we elaborate on the data analysis and interpretation of susceptibilities in the context of $U_c$ that marks the onset of emerging long-range order. Finally, \cref{section:dynamicsus} provides the additional dynamic spectra in the context of magnons and collective charge modes.


\section{Slave-boson mean-field theory \& phase separation}\label{sec:SBMF}

We first provide key results of the established Kotliar-Ruckenstein slave-boson (SB) mean-field (MF) theory \cite{kotliar_new_1986,woelfle_spin_1992,Fresard_1992,woelfle_spin_1997,Hubbard_Wuerzburg,Seufert_2021,Klettthesis,Riegler2022}. Then, we derive a consistent MF treatment of the nearest-neighbor (NN) interaction $V$, which has been introduced in the main text. 

\subsection{Recap of the established MF theory}

The magnetic MF ground state with the ansatz 
\begin{align}\label{eq:mfansatz}
\begin{split}
n_i &\rightarrow \langle n \rangle  = 1 +d^2-e^2 \comma \\
\beta_{0,i} &\rightarrow \langle \beta_0 \rangle =\beta_0 \comma \\
\bs S_i&\rightarrow \langle \bs S_i \rangle=p_0 p\left[ \cos(\QSO \cdot  \bs r_i), \sin(\QSO \cdot \bs r_i),0 \right]^\top \comma \\
\bs \beta_i&\rightarrow \langle \bs \beta_i \rangle =\beta \left[ \cos(\QSO \cdot  \bs r_i), \sin(\QSO \cdot \bs r_i),0 \right]^\top \comma\\
\alpha_i&\rightarrow \langle \alpha \rangle = \alpha \comma
\end{split}
\end{align}
is determined by the saddle point of the Free Energy, i.e., by minimizing wrt.~the SB fields $e,d,p_0,p$ as well as the MF spin ordering vector $\QSO$, while maximizing wrt.~the Lagrange multipliers $\alpha,\beta_0,\beta$. This basis of seven MF variables will be denoted by $\bs\psi$ in the following. Necessary constraints that ensure that (i) every lattice site $i$ is occupied by exactly one boson and (ii) bosonic and fermionic representations of the charge- (iii) and spin-density operators match, are enforced on average by the Lagrange multipliers, yielding 
\begin{equation}\label{eq:MFFE:full}
F (n,T)=  -\frac{T}{N}\sum_{\bk, s = \pm} \ln\left(1+\e^{-\ev_{\bk}^{s}/T}\right) 
+ U d^2 +  \alpha(e^2+d^2+p_{0}^2+p^2-1)
-\beta_0 (2d^2+p_0^2+p^2) -2\beta p_0 p 
+\muz n \period
\end{equation}
Here, $U$ is the on-site Hubbard interaction, $N$ the number of lattice sites, $T$ the temperature, $\muz$ the chemical potential and $n$ the electron filling per site.
The constrains associated with $\alpha$ and $\beta_0$ can be enforced by directly eliminating two arbitrary SB fields and substitution of an effective chemical potential $\mueff$ \cite{Hubbard_Wuerzburg}
\begin{subequations}\label{SBT:MF:Hub:reduction}
	\begin{align}
	1&=e^2+d^2+p_0^2+ p ^2 \label{App:MagneticMF:reduction:alpha}\comma \\
	n &=\langle n \rangle =2d^2+p_0^2+p^2 \label{App:MagneticMF:reduction:n} \comma \\
	\mueff&=\mu_0-\beta_0\label{App:MagneticMF:reduction:mueff} \period
	\end{align}
\end{subequations}
The respective effective Free energy per lattice site and saddle point equations for a minimal set of independent variables where we substituted $e$ and $d$ without loss of generality are given by 
\begin{subequations}\label{SBT:MF:eq:HubSaddle}
	\begin{equation}
	F_{\text{eff}}~\Bigg{|}_{\eqref{SBT:MF:Hub:reduction}}=-\frac{T}{N}\sum_{\bk,s}  \ln\left(1+\e^{-\ev_{\bk,\mueff}^{s}/T}\right)\Bigg{|}_{\eqref{SBT:MF:Hub:reduction}}
	+\frac{U}{2}\left(n-p_0^2-p^2\right)-2\beta p_0 p +\mueff n \comma 
	\end{equation}
	\begin{equation}\label{eq:saddlepointeq}
	\pdd{F_{\text{eff}}}{p}~\Bigg{|}_{\eqref{SBT:MF:Hub:reduction}}=\pdd{F_{\text{eff}}}{p_0}~\Bigg{|}_{\eqref{SBT:MF:Hub:reduction}}=\pdd{F_{\text{eff}}}{\beta}~\Bigg{|}_{\eqref{SBT:MF:Hub:reduction}}
	=\pdd{F_{\text{eff}}}{\mueff}~\Bigg{|}_{\eqref{SBT:MF:Hub:reduction}}	=\pdd{F_{\text{eff}}}{\QSO}~\Bigg{|}_{\eqref{SBT:MF:Hub:reduction}}= 0 \period
	\end{equation}
\end{subequations}
These saddle point equations hold for arbitrary commensurate or incommensurate values of the spin ordering vector, antiferromagnetism is recovered for $\QSO=(\pi,\pi)^\top$.
Thereby, the eigenvalues of the kinetic part of the SB-renormalized Hubbard Hamiltonian are given by \cite{Riegler2022}
\begin{subequations}
	\begin{equation}\label{App:Hopping:eq:EvH}
	\evL_{\bk,\mueff}^{\pm}=\frac12 \left[ \left(\mathcal{Z}^2_+ + \mathcal{Z}^2_-\right)\left(\xi_\bk +\xi_{\bk-\bs Q_{\text{\tiny SO}}} \right) \pm \Delta_{\bk}^{\bs Q_{\text{\tiny SO}}} \right] -\mueff  \comma
	\end{equation}
	with
	\begin{equation}\label{App:Hopping:eq:GapH}
	\Delta_{\bk}^{\QSOs}=\sqrt{\left[\left(\mathcal{Z}^2_+ - \mathcal{Z}^2_-\right)\left(\xi_\bk-\xi_{\bk-\QSOs} \right) \right]^2 + 4\left[ \mathcal{Z}_+\mathcal{Z}_-\left(\xi_\bk +\xi_{\bk-\QSOs} \right) + \beta \right]^2}
	\end{equation}
	representing the magnetic gap. The SB-dependent z-factors are given by
	\begin{equation}\label{App:MagMeanfield:Z_i}
	\mathcal{Z}_\pm
	= \frac{z_+ \pm z_-}{2} \comma \quad 
	z_\pm =\frac{p_0(e+d)\pm p(e-d)}{\sqrt{2\left[1-d^2-(p_0\pm p)^2/2\right] \left[1-e^2-(p_0\mp p)^2/2\right]}} \comma 
	\end{equation}
	and the bare dispersion relation for NN hopping $t$ and NNN hopping $t'$ on the square lattice yields
	\begin{equation}
	\xi_\bk=-2t\left(\cos k_x + \cos k_y \right) -4t' \cos k_x \cos k_y \period
	\end{equation}
\end{subequations}
The MF values of the eliminated variables $e$ and $d$ at the saddle point can directly be recovered with Eqs. \eqref{App:MagneticMF:reduction:alpha} and \eqref{App:MagneticMF:reduction:n}.
In order to recover $\muz$ and $\beta_0$, we need to resubstitute and thereby lift the constraint in \eq{App:MagneticMF:reduction:n}, i.e., express $\langle n\rangle =2d^2+p_0^2+p^2$ as function of the MF variables rather than as being constant, while applying \eq{App:MagneticMF:reduction:alpha} to still substitute the $e$-field. By taking the derivative wrt. $d$ in \eq{eq:MFFE:full}, we the find
\begin{equation}\label{SBT:MF:eq:beta0resubstitute}
\beta_0=\frac{1}{4 d}\frac{1}{N}\sum_{\bs k,s}  \nfd\left(\ev_{\bk,\mueff}^{s}\right)\pdd{\ev_{\bk,\mueff}^{s}}{d}\Bigg{|}_{\eqref{App:MagneticMF:reduction:alpha},\psimf}+ \frac{U}{2}  \comma
\end{equation}
where $\nF$ is the Fermi-Dirac distribution. This derivative has to be evaluated at the saddle point solution, which is denoted by $\psimf$. The MF solution of $\muz$ can now be recovered by utilizing the saddle point results for $\mueff$ and $\beta_0$ with \eq{App:MagneticMF:reduction:mueff}.
Recovering $\muz$ and $\beta_0$ is important in oder to detect phase separation (PS) and perform a consistent fluctuation calculation. The saddle point value for $\alpha$ can be calculated analogously
\begin{equation}
\alpha=-\frac{1}{2  e}\frac{1}{N}\sum_{\bs k,s}  \nfd\left(\ev_{\bk,\mueff}^{s}\right)\pdd{\ev_{\bk,\mueff}^{s}}{e}\atmf 
\comma
\end{equation}
it does not play a role on MF level, but needs to be determined correctly for a consistent fluctuation calculation around the saddle point.

\subsection{Nearest-neighbor interaction $V$ within the MF description}

Within the previously discussed magnetic MF ansatz defined by \eq{eq:mfansatz}, the density $n_i$ is assumed to be uniform, i.e., it does not depend on the lattice site $i$. Therefore, the nearest-neighbor interaction adapts a simple bosonic form
\begin{equation}
H_{V}= V \sum_{<ij>_1} n_i n_j \rightarrow 2 N V   \left( 2d^2+p_0^2+p^2 \right)^2,
\end{equation}
where we applied the SB representation of the density $\langle n\rangle =2d^2+p_0^2+p^2$.

\subsubsection{Invariance of saddle point equations}
Within the Free Energy description, the average filling $\langle n\rangle =2d^2+p_0^2+p^2$ is constrained to the external parameter $n$, which does not depend on the SB fields and is controlled by the chemical potential $\muz$. Therefore, the NN interaction only adds a constant global energy shift, which can rigorously be shown with the previously discussed substitution in \eq{SBT:MF:Hub:reduction}, yielding
\begin{equation}\label{eq:FeffV}
F_{\text{eff}}\rightarrow ~F_{\text{eff}} \Big{|}_{V=0} + 2Vn^2  
\end{equation}
The associated saddle point equations, i.e., \eq{SBT:MF:eq:HubSaddle} are agnostic to that additional term, i.e., the MF solution of the SB fields ($e,d,p_0,p$), the spin ordering vector $\QSO$ as well as Lagrange multiplier $\beta$ and $\mueff$ are independent of $V$. As a consequence the MF band structure and Green's function, which are determined by these quantities, do not depend on $V$.

\subsubsection{Chemical potential shift and phase separation}
However, the saddle point values of the chemical potential $\muz$ and the Lagrange multipliers $\beta_0$ and $\alpha$ do depend on $V$. These can be calculated with a resubstitution in analogy to the previous section, i.e.,
\begin{subequations}
	\begin{align}\label{SBT:MF:eq:beta0resubstitute}
	\beta_0
	&=\phantom{-}\frac{1}{4  d}\frac{1}{N}\sum_{\bs k,s}  \nfd\left(\ev_{\bk,\mueff}^{s}\right)\pdd{\ev_{\bk,\mueff}^{s}}{d}\Bigg{|}_{\eqref{App:MagneticMF:reduction:alpha},\psimf}+ \frac{U}{2} 
	+ \underbrace{\frac{1}{4d} \frac{\partial }{\partial d} 2V \left( 2d^2+p_0^2+p^2 \right)^2 \Bigg{|}_{\psimf}}_{4V n} \comma \\
	\alpha&=
	-\frac{1}{2  e}\frac{1}{N}\sum_{\bs k,s}  \nfd\left(\ev_{\bk,\mueff}^{s}\right)\pdd{\ev_{\bk,\mueff}^{s}}{e}\atmf 
	- \underbrace{\frac{1}{2e} \frac{\partial }{\partial e} 2V \left( 1+d^2-e^2 \right)^2 \Bigg{|}_{\psimf}}_{-4V n} \period \label{eq:alpha:V}
	\end{align}
\end{subequations}
As opposed to $\beta_0$, $\alpha$ has no relevance on MF level, but its consistent treatment is important for the consecutive fluctuation calculation as we will further discuss later on. 
The chemical potential can be recovered with \eq{App:MagneticMF:reduction:mueff}, yielding
\begin{subequations}
	\label{eq:mubeta}
	\begin{align}
	\muz &= \muz \Big{|}_{V=0} + 4Vn  \\
	\beta_0 &= \beta_0 \Big{|}_{V=0} + 4Vn \period
	\end{align}
\end{subequations}
This result is in agreement with the expected behavior that can also be inferred directly from \eq{eq:FeffV}
\begin{equation}
\label{eq:muz}
\muz=\frac{\partial F}{\partial n} = \frac{\partial F_{\text{eff}} \Big{|}_{V=0} + 2Vn^2}{\partial n}
=\muz \Big{|}_{V=0} + 4Vn \period
\end{equation}
Consequently, the chemical potential simply gets an additional contribution proportional to $n$ compared to the result at $V=0$. Since phase separation (PS) occurs if the electronic compressibility $\kappa_T\propto \partial n/\partial \muz$
becomes negative as discussed in the main text, we recognize that positive $V$ inhibits PS and leads to its complete elimination for sufficiently high $V$.
PS requires two energy-degenerate solutions of the grand potential $\Omega(\muz,T)$ with different fillings at the same critical chemical potential $\muz^c$. In order to lift this degeneracy one has to introduce a filling-depended term in the Lagrangian. A term that scales linearly with $n$ can be absorbed in the effective chemical potential and just shifts the map of filling and $\mueff$ by a constant factor like an on-site potential. Therefore the simplest density-dependent term that inhibits phase separation scales with $n^2$, which happens to be the nearest-neighbor density-density interaction in our MF picture.  

\section{Fluctuations around a symmetry-broken ground state}\label{sec:SBfluc}

First, we recap the SB fluctuation formalism \cite{woelfle_spin_1997,Hubbard_Wuerzburg} that has recently been extended to encompass magnetic saddle points \cite{Seufert_2021}. Then, we derive additional contributions to the fluctuation matrix that arise due to the newly-implemented interaction $V$ with the final result being \eq{eq:MV:result}.
Around the (magnetic) MF saddle point, the leading order of fluctuations is quadratic and described by the fluctuation matrix

\begin{equation}\label{eq:flucmatrixdef}
\mathcal{M}_{\mu\nu}^{ab}(\bs q,\i\omega_n)
=\frac12 \frac{\delta^2 \mathcal{S}(\psi)}{\delta \psi_{\mu,-\bs q - a\QSO, -\i\omega_n} \, \delta \psi_{\nu,\bs q+b\QSO, \i\omega_n}} \comma
\end{equation}
where the bosonic fluctuation field $\delta \psi_\mu =\psi_\mu - \langle \psi_\mu \rangle$ defines the deviation from the MF value $\langle \psi_\mu \rangle$. For fluctuations around AFM mean-field saddle points, i.e. $\QSO=(\pi,\pi)^\top$, we have $a,b \in \{0,1\}$ such that the dimension of the basis is doubled compared to the homogeneous PM or FM case.
Moreover, we define the fluctuation field Green's function by means of the inverse of the fluctuation matrix
\begin{equation}
\mathcal{G}^{ab}_{\mu\nu}(\bs q,\i\omega_n) = - \left[  \mathcal{M}^{-1}(\bs q,\i\omega_n) \right]_{\mu\nu}^{ab} \period
\end{equation}
The fluctuation fields are complex in general, however, we adapt the established radial gauge, where only the $d$-field remains to be complex, i.e., $d=d_1+\i d_2$, while the phases of the other fields are gauged away. In Refs. \cite{Hubbard_Wuerzburg,Seufert_2021}, it has been discussed that an arbitrary field can be eliminated from the fluctuation basis with application of \eq{App:MagneticMF:reduction:alpha} in analogy to the MF calculation. Thereby, it has been shown that $\mathcal{G}$ (and therefore observables like susceptibilities) does not depend on whether this substitution has been applied or which field has been substituted, if done consistently. Therefore, it is usually recommended to apply the substitution to save calculation time in the numerical analysis. However, here we choose to keep all fields in our fluctuation basis because it makes the implementation of the nearest-neighbor interaction $V$ more convenient, i.e.,
\begin{equation}\label{eq:flucbasis}
\bs \psi = \left(\alpha,e,d_1,d_2,p_0,\beta_0,p_1,\beta_1,p_2,\beta_2,p_3,\beta_3 \right)^\top\period
\end{equation}

\subsection{Recap of the established fluctuation formalism}

Here, we provide results for the fluctuation matrix around AFM mean-field saddle points. Moreover, we recap, how susceptibilities are calculated by means of inverse fluctuation matrix elements.

\subsubsection{Fluctuation matrix around AFM saddle points}

The fluctuation matrix is composed of a bosonic part that results from the coupling of bosons to the local Hubbard interaction $U$ and Lagrange multipliers $\alpha,\beta_0,\bs \beta$ and a non-local, pseudofermionic part that couples to the MF Green's function within a diagrammatic expansion
\begin{equation}
\mathcal{M}=\mathcal{M}^B+\mathcal{M}^F \period
\end{equation}
The bosonic part is calculated by Eq.~B.14 in Ref.~\cite{Seufert_2021}. For the basis defined in \eq{eq:flucbasis} and fluctuations around an AFM mean-field ground state, i.e., $\QSO=(\pi,\pi)^\top$, the result is given by
\allowdisplaybreaks
\begin{subequations}\label{eq:Mbosonicresult}
	\begin{align}
	\left[\mathcal{M}^B\right]^{0,0}
	=\left[\mathcal{M}^B\right]^{1,1} 
	&= 
	\begin{pmatrix}
	0 & e & d & 0 & p_0 & 0 & 0 & 0 & 0 & 0 & 0 & 0 \\ 
	e & \alpha & 0 & 0 & 0 & 0 & 0 & 0 & 0 & 0 & 0 & 0 \\
	d & 0 & U-2\beta_0 + \alpha &  \omega_n & 0 & -2 d & 0 & 0 & 0 & 0 & 0 & 0 \\
	0 & 0 & -\omega_n & U-2\beta_0 +\alpha & 0 & 0 & 0 & 0 & 0 & 0 & 0 & 0 \\
	p_0 & 0 & 0 & 0 & -\beta_0 +\alpha& - p_0 & 0 & 0 & 0 & 0 & 0 & 0 \\
	0 & 0 & -2 d & 0 & - p_0 & 0 & 0 & 0 & 0 & 0 & 0 & 0 \\
	0 & 0 & 0 & 0 & 0 & 0 & -\beta_0 +\alpha& - p_0 & 0 & 0 & 0 & 0 \\
	0 & 0 & 0 & 0 & 0 & 0 & - p_0 & 0 & 0 & 0 & 0 & 0 \\
	0 & 0 & 0 & 0 & 0 & 0 & 0 & 0 & -\beta_0 +\alpha& - p_0 & 0 & 0 \\
	0 & 0 & 0 & 0 & 0 & 0 & 0 & 0 & - p_0 & 0 & 0 & 0 \\
	0 & 0 & 0 & 0 & 0 & 0 & 0 & 0 & 0 & 0 & -\beta_0 +\alpha& - p_0 \\
	0 & 0 & 0 & 0 & 0 & 0 & 0 & 0 & 0 & 0 & - p_0 & 0 
	\end{pmatrix}\comma 
	\\
	\left[\mathcal{M}^B\right]^{0,1} =\left[\mathcal{M}^B\right]^{1,0}
	&=
	\begin{pmatrix}
	0 & 0 & 0 & 0 & 0 & 0 & p & 0 & 0 & 0 & 0 & 0 \\
	0 & 0 &0 & 0 & 0 & 0 & 0 & 0 & 0 & 0 & 0 & 0 \\
	0 & 0 & 0 & 0 & 0 & 0 & 0 & 0 & 0 & 0 & 0 & 0 \\
	0 & 0 &0 & 0 & 0 & 0 & 0 & 0 & 0 & 0 & 0 & 0 \\
	0 & 0 &	0 & 0 & 0 & 0 & - \beta & - p & 0 & 0 & 0 & 0 \\
	0 & 0 &	0 & 0 & 0 & 0 & - p & 0 & 0 & 0 & 0 & 0 \\
	p & 0 &	0 & 0 & -\beta & - p & 0 & 0 & 0 & 0 & 0 & 0 \\
	0 & 0 &	0 & 0 & - p & 0 & 0 & 0 & 0 & 0 & 0 & 0 \\
	0 & 0 &	0 & 0 & 0 & 0 & 0 & 0 & 0 & 0 & 0 & 0 \\
	0 & 0 &	0 & 0 & 0 & 0 & 0 & 0 & 0 & 0 & 0 & 0 \\
	0 & 0 &	0 & 0 & 0 & 0 & 0 & 0 & 0 & 0 & 0 & 0 \\
	0 & 0 &	0 & 0 & 0 & 0 & 0 & 0 & 0 & 0 & 0 & 0 
	\end{pmatrix} \period
	\end{align}
\end{subequations}
The pseudofermionic part is given by
\begin{align}\label{APP:Magfluc:eq:MF}
\begin{split}
\Big[\mathcal{M}^{F} (\bs q, \i\omega_n)\Big]_{\mu \nu}^{a b} 
&=
\frac{1}{2} 
\sum_{\bk}
\sum_s \nF\left(\ev_{\bk}^s\right) 
~\Bigg{[} 
\ul U_{\bs{k}}^{\nodag} \bigg( 
\left[\szzb_{\mu\nu}^\dagger \right]^b
\otimes \left[\underline{\mathcal{H}}_{\bs{k}+a\bs{Q}} \szzb_{0}^a \right] 
+ 
\left[ \szzb^\dagger_0 \right]^b
\otimes
\left[\underline{\mathcal{H}}_{\bs{k} + b\bs{Q}}
\szzb_{\mu\nu}^a
\right]  
\\
&\hspace{75pt} 
+
\left[	\szzb_\mu^\dagger \right]^b
\otimes 
\left[\underline{\mathcal{H}}_{\bs{k} + \bs{q} + (a+b) \bs{Q}} 
\szzb^a_\nu \right]
+
\left[	\szzb_\nu^\dagger \right]^b
\otimes
\left[\underline{\mathcal{H}}_{\bs{k} - \bs{q}}
\szzb_\mu^a
\right]
\bigg) \ul U_{\bs{k}}^\dagger \Bigg]^{s,s}	
\\[5pt]
\hphantom{=}&+ 
\frac{1}{2} \sum_{\bk} \sum_{s,s'}
\frac{\nF\left(\ev_{\bs{k}}^s\right) -\nF\left(\ev_{\bs{k}+\bs{q}}^{s'}\right)}{\ev_{\bs{k}}^s- \ev_{\bs{k}+\bs{q}}^{s'}+\i\omega_n} \\
&\phantom{==========}\times
\Bigg[	\sum_v\ul U_{\bs{k}}^{\nodag} 
\left(
\left[ \szzb_\mu^\dagger \right]^v
\otimes
\left[\underline{\mathcal{H}}_{\bs{k} + \bs{q} + (v+a) \bs{Q}}
\szzb_0^{v+a}
\right]
+ 
\left[	\szzb_0^\dagger \right]^v
\otimes 
\left[\underline{\mathcal{H}}_{\bs{k}+v\bQ}
\szzb_\mu^{v+a} \right]
+ \ul{\mathfrak{B}}_{\mu}^{v,v+a} 
\right)
\ul U^\dagger_{\bs{k}+\bs{q}} \Bigg]^{s,s'} \\
&\phantom{==========}\times
~\Bigg[\sum_u
\ul U_{\bs{k}+\bs{q}}^{\nodag} 
\left(
\left[ \szzb_\nu^\dagger \right]^{u+b}
\otimes 
\left[\underline{\mathcal{H}}_{\bs{k} + u \bs{Q}} 
\szzb_0^u
\right]
+ 
\left[\szzb_0^\dagger \right]^{u+b} 
\otimes
(\underline{\mathcal{H}}_{\bs{k} + \bs{q} + (u+b) \bs{Q}} 
\szzb_\nu^u
+ \ul{\mathfrak{ B}}_{\nu}^{u+b, u} 
\right) \ul U^\dagger_{\bs{k}}
\Bigg]^{s',s}\comma
\end{split}
\end{align}
where $\ev^s_\bk$ and $\underline{U}_\bk$ are the eigenvalues and matrix of eigenvectors of the MF Hamiltonian respectively, $\mathcal{H}$ is the bare hopping matrix, and $\underline{\bs Z}$ are matrices that depend on the SB z-factors, which are precisely defined in Refs. \cite{Seufert_2021,Riegler2022}.

\subsubsection{Slave-boson susceptibilities}

The charge susceptibility $\chi_c$, longitudinal spin susceptibility $\chi_s^l$ (parallel to MF magnetization) and transversal spin susceptibility $\chi_s^t$ (perpendicular to MF magnetization) are given by superpositions of matrix elements of the fluctuation field Green's function \cite{Seufert_2021,Klettthesis,Riegler2022}
\begin{align} 		\label{eq:suscepts}
\begin{split}
\chi_c(q) &= \langle \delta n_q \delta n_{-q}\rangle = -4  d_1^2 
\mathcal{G}^{0,0}_{d_1,d_1}
-   p_0^2 
\mathcal{G}^{0,0}_{p_0,p_0}
- 2  d_1  p_0
\Big(
\mathcal{G}^{0,0}_{d_1,p_0}
+  \mathcal{G}^{0,0}_{p_0,d_1}
\Big)  
\vphantom{G_{\Big{|}}}
- 2  d_1  p \left(
\mathcal{G}^{0,1}_{d_1,p_1}
+ 
\mathcal{G}^{1,0}_{p_1,d_1}
\right) 
-  p_0  p \left(
\mathcal{G}^{0,1}_{p_0,p_1}
+ 
\mathcal{G}^{1,0}_{p_1,p_0}
\right) 
-  p^2 
\mathcal{G}^{1,1}_{p_1,p_1}\comma\\
\vphantom{\Bigg{|}} \chi_s^l(q)
&=
\langle \delta S^x_q \delta S^x_{-q}\rangle=
- p_0^2 \mathcal{G}^{0,0}_{p_1,p_1}
-   p_0 p \left( 
\mathcal{G}^{0,1}_{p_1,p_0}+\mathcal{G}^{1,0}_{p_0,p_1}
\right)
- p^2 \mathcal{G}^{1,1}_{p_0,p_0}	\comma \\
\vphantom{\Big{|}}\chi_s^t(q)
&=
\langle \delta S^y_q \delta S^y_{-q}\rangle= \langle \delta S^z_q \delta S^z_{-q}\rangle=
- p_0^2 \mathcal{G}^{0,0}_{p_2,p_2}
=
- p_0^2 \mathcal{G}^{0,0}_{p_3,p_3} \comma
\end{split}
\end{align}
where we define $q=(\bq,\i\omega_n)$. Finally, the bare susceptibility of the renormalized MF band structure with fermionic MF Green's function $G^{\langle \bs\psi\rangle}_k$ is given by 
\begin{equation}\label{SM:eq:X0}
\chi_0(q)
= \langle n_q n_{-q} \rangle 
= -\frac{T}{N}\sum_q \tr G^{\langle \bs\psi\rangle}_{k+q} G^{\langle \bs\psi\rangle}_k
=-2\mathcal{M}^{0,0}_{\beta_0,\beta_0} \period
\end{equation}
In contrast to the previously defined susceptibilities, it does not contain bosonic fluctuations and thus never diverges on the basis of a MF saddle point solution. Nevertheless, it is suitable to analyze the fermiology of the MF ground state.

\subsection{Nearest-neighbor interaction $V$ within the fluctuation formalism}

In order to consider fluctuations of the NN interaction $V$
\begin{equation}
H_{V}= V \sum_{<ij>_1} n_i n_j = \frac{V}{2} \sum_{i,\Delta} n_i n_{i+\Delta}
\end{equation}
around the saddle point, we express $n_i$ as function of the SB-fields. We introduce the factor $\frac12$ to avoid double counting and $\sum_\Delta$ denotes the sum over all nearest neighbors. Within the constrained subspace, there are two equivalent descriptions $n_i = 1+d^\dagger_i d^\nodag_i - e^\dagger_i e^\nodag_i = 2 d^\dagger_i d^\nodag_i+ p^\dagger_{0,i}p^\nodag_{0,i}+ p^\dagger_{1,i}p^\nodag_{1,i}+ p^\dagger_{2,i}p^\nodag_{2,i}+ p^\dagger_{3,i}p^\nodag_{3,i}$. 
We choose the former in the following without loss of generality. This choice is more convenient because it only implicates direct couplings between the charge-fields $e_i$ and $d_i$. 
Since such a purely bosonic representation exists, the NN interaction only contributes to the bosonic part of the fluctuation matrix $\mathcal{M}^B$, while $\mathcal{M}^F$ is agnostic to $V$. The MF saddle point value of the Lagrange multiplier $\alpha$ depends on the representation of the operator $n_i$ and needs to be determined consistently according to \eq{eq:alpha:V}, i.e.,
\begin{equation}
\label{eq:alpha}
\alpha = \alpha\Big{|}_{V=0}+4Vn \period
\end{equation}

To calculate the contributions to the fluctuation matrix, we need to expand the interaction Hamiltonian $H_{V}$ around the MF saddle point
\begin{equation}
\delta H_{V} = \frac{V}{2} \sum_{i,\Delta} (n+\delta n_i) (n+\delta n_{i+\Delta}) 
= 2 VN  n^2 + \frac{V}{2} n \sum_{i,\Delta} (\delta n_i + \delta n_{i+\Delta})
+ \frac{V}{2}\sum_{i,\Delta} \delta n_i  \delta n_{i+\Delta} \, ,
\end{equation}
with 
\begin{equation}
\delta n_i 
=  n_i -\langle n \rangle 
= \sum_\mu \frac{\partial n_i}{\partial \psi_{\mu,i}}\atmf \delta \psi_{\mu,i} 
+ \sum_{\mu\nu}\frac{\partial^2 n_i}{\partial \psi_{\mu,\i}  \partial \psi_{\nu,i} }\atmf
\delta \psi_{i,\mu} \delta \psi_{i,\nu} + \mathcal{O}(\delta \psi^3) 
=   \delta n^{(1)} +\delta n^{(2)} + \mathcal{O}(\delta n^{3})
\end{equation}
and collect all second-order terms in fluctuation fields $\delta \psi$ . Thereby, we encounter the on-site term
\begin{equation}
\delta H_{V}^{ii} =\frac{V}{2} n \sum_{i,\Delta} \left(\delta n_i^{(2)} + \delta n^{(2)}_{i+\Delta}\right) = \frac{V}{2} n \sum_{i,\Delta} \delta n_i^{(2)} + \frac{V}{2} n \sum_{i-\Delta,\Delta} \delta n_{i}^{(2)} =4 n V \sum_i \delta n_i^{(2)} \comma
\end{equation}
associated with a second derivative of the charge density and a nearest-neighbor contribution
\begin{equation}
\delta H_{V}^{<ij>_1}=	\frac{V}{2}\sum_{i,\Delta} \delta n_i  \delta n_{i+\Delta} \comma
\end{equation}
associated with a product of two first derivatives. Collecting all second-order terms yields
\begin{equation}
\delta H_{V}^{(2)} = V \sum_i \left(4 n   \delta n^{(2)}_i + \frac12\sum_\Delta  \delta n^{(1)}_i  \delta n^{(1)}_{i+\Delta}  \right) \period 
\end{equation}
In order to calculate contributions to the fluctuation matrix defined in \eq{eq:flucmatrixdef}, we need to expand the SB fields in momentum space. 

\subsubsection{Product of first derivatives}

\begin{equation}
\sum_{i,\Delta} \delta n_i \delta n_{i+\Delta} = \sum_\bk \delta n_{-\bk} \delta n_\bk f^1_\bk
\end{equation}
with
\begin{equation}
f^1_\bk = 2\left[\cos(k_x)+\cos(k_y)\right] \period
\end{equation}
Exploiting that the MF charge density $\langle n \rangle =1+d_1^2+d_2^2-e^2$ is spatially uniform, we find
\begin{equation}
\delta n_\bk^{(1)} = \sum_{\tilde{\mu}} n_{\tilde{\mu}} \delta \psi_{\tilde{\mu},\bk}\comma
\end{equation}
with $n_\mu = \partial \langle n \rangle/\partial \psi_\mu$. 
For more details on this momentum-space expansion of fluctuation fields see, e.g., Eq. A40 in Ref.~\cite{Riegler2022}.
Note, that the MF charge density is a real quantity, i.e., $\langle d_2 \rangle=0$. Combining all previous results yields
\begin{equation}
\left[\mathcal{M}_V^{<ij>_1}(\bs q)\right]_{\mu\nu}^{ab}=
\frac12 
\frac{
	\delta^2\mathcal{S}^{\langle ij\rangle}_V
}
{\delta \psi_{\mu,-\bs q - a\QSO, -\i\omega_n} \, \delta \psi_{\nu,\bs q+b\QSO, \i\omega_n}}
=
\frac12 
\frac{
	\frac{V}{2}\sum_\bk\sum_{\i\tilde{\omega}_n}\sum_{\tilde{\mu}} n_{\tilde{\mu}} \delta \psi_{\tilde{\mu},-\bk,-\i\tilde{\omega}_n}\sum_{\tilde{\nu}} n_{\tilde{\nu}} \delta \psi_{\tilde{\nu},\bk,\i\tilde{\omega}_n} f^1_\bk
}
{\delta \psi_{\mu,-\bs q - a\QSO, -\i\omega_n} \, \delta \psi_{\nu,\bs q+b\QSO, \i\omega_n}}
= \frac{V}{2} n_{\mu} n_{\nu} f^1_{\bq+a\QSO}\delta_{a,b}
\end{equation}
as momentum-dependent contribution to the fluctuation matrix.

\subsubsection{Second derivative}

Once more exploiting that the MF charge density is uniform, we find
\begin{equation}
\delta n^{(2)}_\bk =
\sum_{\tilde{\mu},\tilde{\nu},\bk_1} n_{\tilde{\mu}\tilde{\nu}}
\delta \psi_{\bk_1,\tilde{\mu}} \delta \psi_{\bk-\bk_1\tilde{\nu}}\comma
\end{equation} 
with $n_{\mu\nu} = \partial^2 \langle n \rangle/\partial \psi_\mu\partial\psi_\nu$. Since the term is on-site in real space, we are projected onto the $\bk=0$ channel, yielding
\begin{align}
\left[\mathcal{M}_V^{<ii>}\right]_{\mu\nu}^{ab}=
\frac12 
\frac{
	\delta^2\mathcal{S}^{\langle ii\rangle}_V
}
{\delta \psi_{\mu,-\bs q - a\QSO, -\i\omega_n} \, \delta \psi_{\nu,\bs q+b\QSO, \i\omega_n}}
=\frac12
\frac{
	4Vn\sum_{\bk_1}\sum_{\i\tilde{\omega}_n}\sum_{\tilde{\mu},\tilde{\nu}} n_{\tilde{\mu},\tilde{\nu}} \delta \psi_{\tilde{\mu},\bk_1,\i\tilde{\omega}_n} \delta \psi_{\tilde{\nu},0-\bk_1,-\i\tilde{\omega}_n}
}
{\delta \psi_{\mu,-\bs q - a\QSO, -\i\omega_n} \, \delta \psi_{\nu,\bs q+b\QSO, \i\omega_n}}
= 2 V n n_{\mu,\nu} \delta_{a,b}
\end{align}
as constant contribution to the fluctuation matrix.

\subsubsection{Final result of the fluctuation matrix}

We are left to calculate the derivatives of $\langle n\rangle$ and to add up both terms. Additional matrix elements only arise in association with the fields $e,d_1,d_2$. All non-zero contributions are therefore given by
\begin{subequations}\label{eq:MV:result}
	\begin{align}
	\left[\mathcal{M}^B_V(\bq)\right]^{0,0} 
	&=V
	\begin{pmatrix}
	-4n+2e^2 f^1_\bq & -2ed f^1_\bq & 0 \\
	-2edf^1_\bq & 4n+2d^2 f^1_\bq & 0 \\
	0 & 0 & 4n  \comma  
	\end{pmatrix}
	\comma \quad
	\bs \psi=\begin{pmatrix}
	e\\d_1\\d_2
	\end{pmatrix}\\
	\left[\mathcal{M}^B_V(\bq)\right]^{1,1} &=V
	\begin{pmatrix}
	-4n+2e^2 f^1_{\bq+\QSO} & -2ed f^1_{\bq+\QSO} & 0 \\
	-2ed f^1_{\bq+\QSO} & 4n+2d^2 f^1_{\bq+\QSO} & 0 \\
	0 & 0 & 4n
	\end{pmatrix}
	\comma \quad
	\bs \psi=\begin{pmatrix}
	e\\d_1\\d_2
	\end{pmatrix}\\
	\left[\mathcal{M}^B_V(\bq)\right]^{0,1} &=	\left[\mathcal{M}^B_V(\bq)\right]^{1,0}=0\period 
	\end{align}
\end{subequations}
These extra matrix elements need to be added to the results of Eqs.~\eqref{eq:Mbosonicresult} and \eqref{APP:Magfluc:eq:MF} to obtain the full fluctuation matrix that describes fluctuations around AFM saddle points that are subject to the interactions $U$ and $V$. 

\subsection{Longer ranged density-density interaction}

Our results can easily be generalized to encompass longer-ranged interactions like, e.g., a NNN interaction $V_2$ etc. 
\begin{equation}
H_{V}= V \sum_{<ij>_1} n_i n_j = \frac{V}{2} \sum_{i,\Delta} n_i n_{i+\Delta} \quad \rightarrow \quad H_{V}= \sum_{l,<ij>_l} V_l \,  n_i n_j = \sum_{l,i,\Delta_l} \frac{V_l}{2} \, n_i n_{i+\Delta_l} \period
\end{equation}
\eq{eq:mubeta} reads now
\begin{subequations}
	\begin{align}
	\muz &= \muz \Big{|}_{V=0} +  n  \,\sum_l Z^l V_l \comma \\
	\beta_0 &= \beta_0 \Big{|}_{V=0} + n \,\sum_l  Z^l V_l\, ,
	\end{align}
	with analogous changes for alpha 
	\begin{align}
	\alpha &= \alpha \Big{|}_{V=0} + n\,\sum_l Z^l V_l \comma
	\end{align}
\end{subequations}
where $Z^l$ is the number of l-th nearest neighbors.
Respective contributions of the fluctuation formalism simply yield additional form-factors $f^l_\bk$, with 
\begin{equation}
\begin{aligned}
f^1_\bk &= 2 \cos(k_x)+ 2 \cos(k_y), \\
f^2_\bk &= 4 \cos(k_x) \cos(k_y), \\
f^3_\bk &= 2 \cos(2k_x) + 2 \cos(2k_y), \\
f^4_\bk &= 4 \cos(2k_x) \cos(k_y) + 4 \cos(k_x)\cos(2k_y), \\
\vdots
\end{aligned}
\end{equation}
and equation \eqref{eq:MV:result} taking the form
\begin{subequations}
	\begin{align}
	\left[\mathcal{M}^B_V(\bq)\right]^{0,0} 
	&=\sum_l V_l \,
	\begin{pmatrix}
	-Z^l n+2 e^2 f^l_\bq & -2 ed f^l_\bq & 0 \\
	-2 ed f^l_\bq &  Z^ln+2 d^2 f^l_\bq & 0 \\
	0 & 0 &  Z_ln  \comma  
	\end{pmatrix}
	\comma \quad
	\bs \psi=\begin{pmatrix}
	e\\d_1\\d_2
	\end{pmatrix}\\
	\left[\mathcal{M}^B_V(\bq)\right]^{1,1} &=\sum_l V_l \,
	\begin{pmatrix}
	- Z^ln+2 e^2 f^l_{\bq+\QSO} & -2 ed f^l_{\bq+\QSO} & 0 \\
	-2 ed f^l_{\bq+\QSO} &  Z_l n+2 d^2 f^l_{\bq+\QSO} & 0 \\
	0 & 0 & Z^ln
	\end{pmatrix}
	\comma \quad
	\bs \psi=\begin{pmatrix}
	e\\d_1\\d_2
	\end{pmatrix}\\
	\left[\mathcal{M}^B_V(\bq)\right]^{0,1} &=	\left[\mathcal{M}^B_V(\bq)\right]^{1,0}=0\period 
	\end{align}
\end{subequations}

\section{Maxwell constructions}\label{sec:maxwell}

\begin{figure}[h!]
	\includegraphics[width=0.995\textwidth]{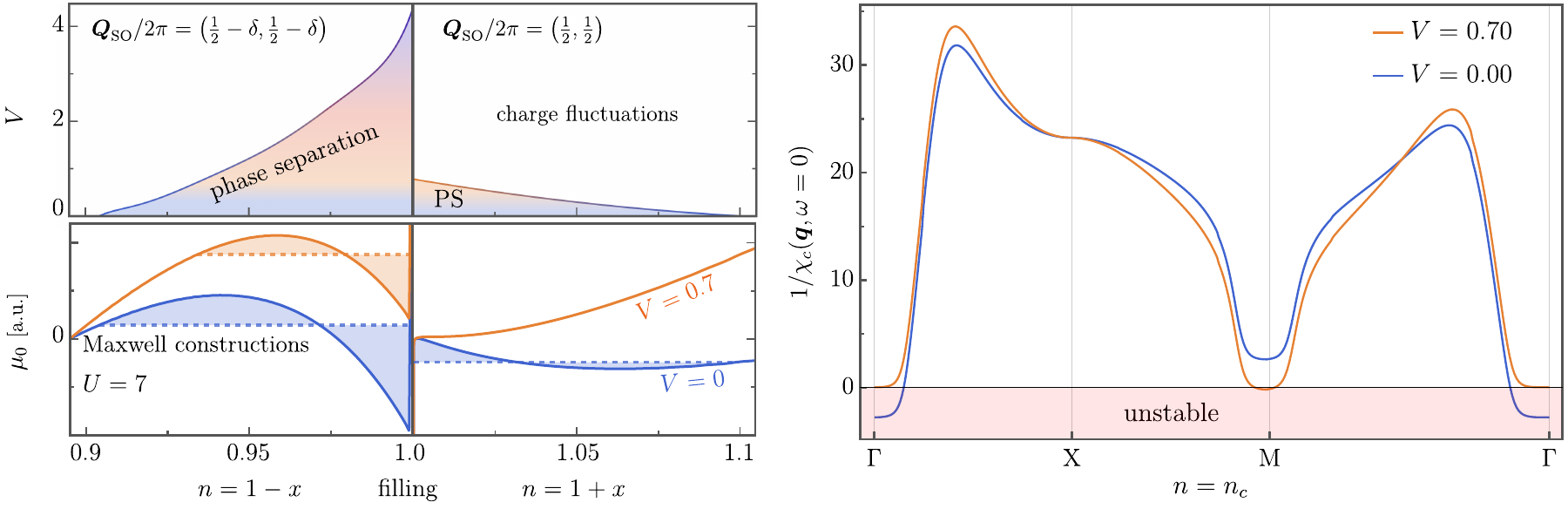}
	\caption{Left panel: Analysis of mean-field phase separation (PS) for $U=7$ as function of the of the electron filling $n$ and NN interaction $V$ at zero temperature. PS domain in the $V-n$ plane on top and chemical potential $\mu_0(n)$ with exemplarily Maxwell constructions on the bottom. Right panel: Static charge susceptibility on the high symmetry path of the BZ at the onset of MF PS, i.e. $n=n_c=1.006, V=V_c=0.7$ along with $V=0$ for comparison. Mean-field PS is accompanied by a divergence of $\chi_c$ at the $\Gamma$-point that disappears for $V>V_c$ as expected.}
	\label{fig:PSSM}
\end{figure}

Phase separation (PS) is characterized by a breakdown of the bijective Legendre transformation between the Free Energy $F(n,T)$ and the Grand Potential $\Omega(\mu_0,T)$, accompanied by a negative compressibility $\kappa_T \propto \partial n/\partial \mu_0 < 0$. The left panel of \autoref{fig:PSSM} shows the phase separated (PS) domains in the $V-n$ plane exemplarily for the Hubbard repulsion $U=7$.
The doping-range that is affected by PS can be recovered by means of Maxwell constructions, where the areas enclosed by the function $\mu_0(n)$ above and blow $\mu_0^c$ are equal in size as shown in the bottom panel for two specific values of $V$. Thereby, $\mu_0^c$ marks the critical chemical potential that hosts two energy-degenerate ground states of the grand potential $\Omega(\mu_0,T)$ with different electron fillings. At $V=0$, PS is present for both, electron- and hole-doping with $x \lesssim 0.1$. However, the enclosed area of the Maxwell construction -- which is proportional to the condensation energy of the PS state, see e.g., \cite{Klettthesis,Riegler2022} -- is much larger for hole-doping, indicating a higher robustness of the corresponding PS domain. With the addition of $V$, the chemical potential is affected according to \eq{eq:mubeta}, which leads to a gradual reduction of PS to smaller dopings, until it is completely removed at a critical value $V_c$ where $\mu_0(n)$ becomes monotonous as demonstrated in \autoref{fig:PSSM}. 
At $V_c=0.7$, $\mu_0(n)$ features a saddle point at a critical filling $n_c=1.006$, which implicates a divergence of the compressibility $\kappa_T^{-1}=0$. That singularity has to be accompanied by a divergence of the static charge susceptibility at the $\Gamma$-point, which is present in our theory as show in the right panel of \autoref{fig:PSSM}. 
For $V<V_c$, the charge susceptibility becomes negative around the $\Gamma$-point, which reflects the instability of the homogeneous MF parent state at parameter domains that are deeper within the PS region. These findings underline the consistency between the MF and fluctuation formalism in regard of finite $V$, that was already established in Ref.~\cite{Seufert_2021} for $V=0$.

\section{Mean-field fermiology: Hole-doping vs. electron-doping } \label{SM:sec:fermiology}

\begin{figure}[h!]
	\includegraphics[width=0.99\linewidth]{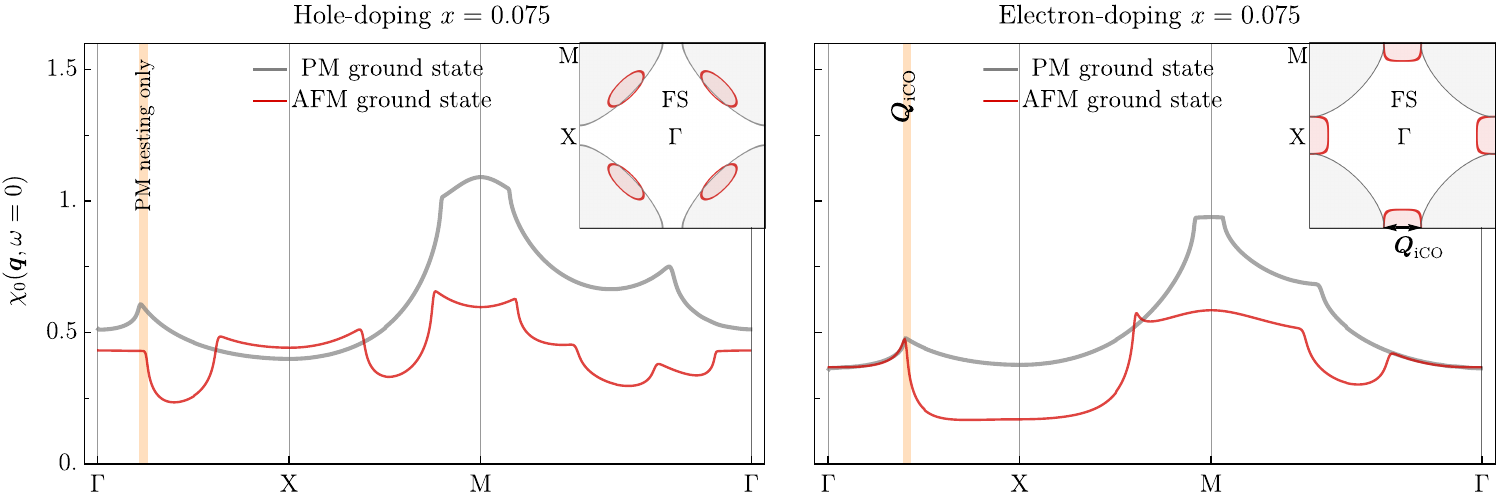}
	\caption{Bare susceptibility $\chi_0$ on the high symmetry path of the BZ for $x=0.075$, comparing electron- and hole-doping, based on a paramagnetic (PM) mean-field ground states at $U=2.25$ and AFM saddle points for $U=6$. The insets show respective Fermi surfaces for the AFM in red and PM in gray.}\label{SM:fig:X0}
\end{figure} 
The bare susceptibility $\chi_0$ defined in \eq{SM:eq:X0} does not encompass bosonic fluctuations and thus does not feature divergences but is suitable to describe the fermiology of the renormalized (magnetic) mean-field ground state. $\chi_0$ does not depend on $V$ since the MF band structure is agnostic to $V$ and it becomes equivalent to the charge susceptibility in the zero-interaction limit ($\chi_0\rightarrow \chi_c$ for $U\rightarrow 0$ \cite{Hubbard_Wuerzburg}). A peak in $\chi_0$ implicates Fermi-surface (FS) nesting, which can lead to a divergence in $\chi_c$ at sufficiently high $U$, like in the case of emerging CO at $\QCO$ for electron-doping. \autoref{SM:fig:X0} compares $\chi_0$ for hole- and electron-doping $x=0.075$ based on PM and AFM mean-field ground states respectively and displays associated Fermi surfaces as an inset. For electron-doping there is a nesting peak at the same wave vector $\QCO$ within PM and AFM ground states. Charge-correlations are, therefore, expected in both, PM and AFM states as a rather universal feature.
In contrast, for hole-doping the AFM transition inhibits nesting, implying that nesting-enhanced CO should not be present here within the magnetic regime.

\section{Static susceptibilities \& critical interaction $U_c$}\label{sb:sec:uc}

\begin{figure}[b!]
	\includegraphics[width=0.999\linewidth]{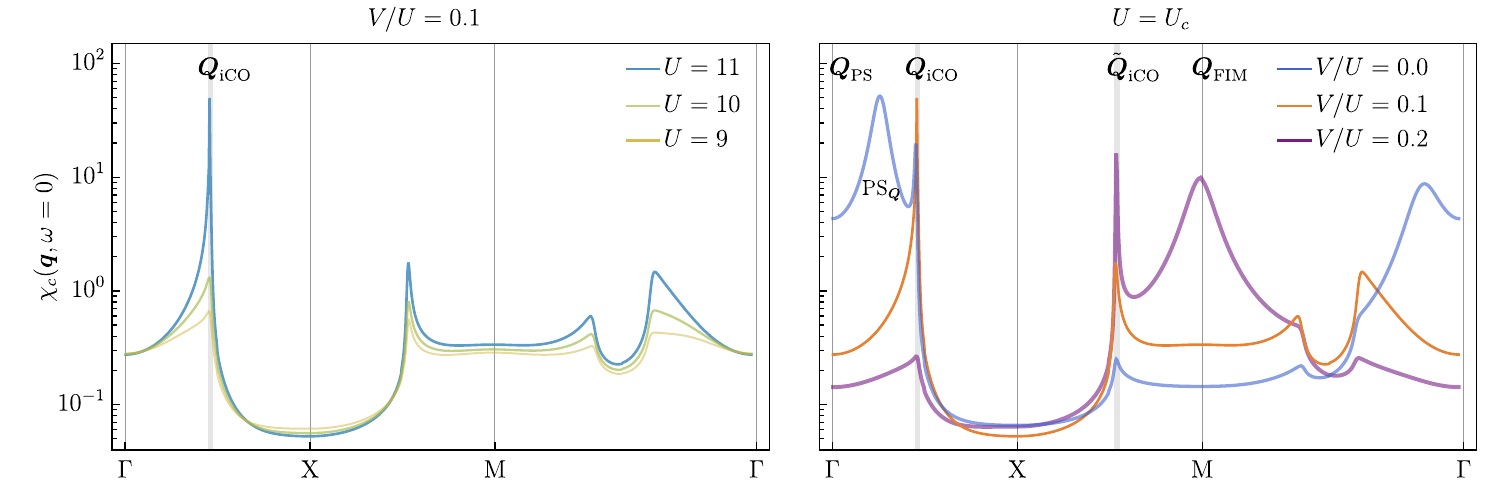}
	\caption{Static charge susceptibility $\chi_c (\bs q, \omega = 0)$ on the high symmetry path of the BZ (with respect of the PM reciprocal lattice vectors) for $n=1.1$ and different interactions $U,V$ at low temperature $T=0.005$.}\label{SM:Fig:chi}
\end{figure}

Here, we elaborate on the data analysis that was performed to create Figure 2 in the main text and provide further background information regarding the behavior of susceptibilities and the critical interaction $U_c$ that marks the onset of long-range order. 
Within the stable AFM domain that is present for electron-doping at sufficiently low interaction, the static charge susceptibility $\chi_c(\bq,\omega=0)$ and longitudinal spin susceptibility $\chi_s^l(\bq,\omega=0)$ remain finite, whereas the transversal spin susceptibility $\chi_s^t(\bq,\omega=0)$ features a divergence at $\QSO=(\pi,\pi)$ due to the Goldstone mode associated with the symmetry-broken AFM mean-field ground state \cite{Seufert_2021}. The definition of the aforementioned susceptibilities and how to calculate them via SB fluctuations has been provided in \eq{eq:suscepts}. Within the SB formalism, we can directly address the static limit $\omega=\omega_n=0$ where the susceptibilities are real-valued, i.e., without addition of a convergence parameter. Thereby, we achieve good convergence for $N= 512\times512$ lattice sites in our analysis.

\begin{figure}[t!]
	\includegraphics[width=0.99\linewidth]{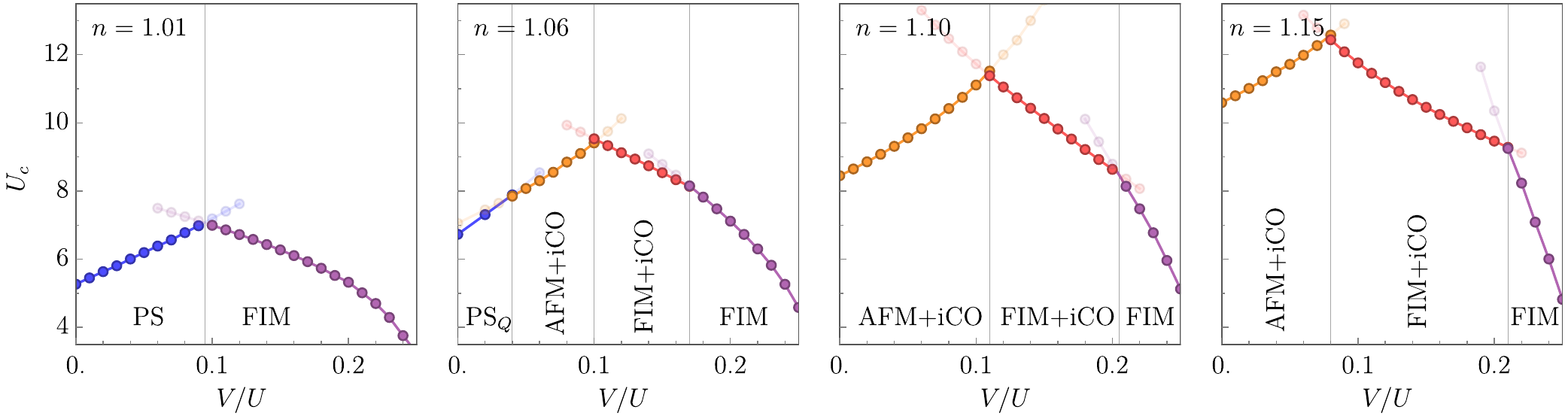}
	\caption{Critical interaction $U_c$ that marks the onset of long-range order, implicated by a divergence of the static charge susceptibility $1/\chi_c(\bq=\qco, \omega = 0)=0$, as function of the interaction ratio $V/U$ for four different fillings.
		The respective ordering vector $\qco$ dictates the type of emerging charge-order, i.e., phase-separation (PS) for $\qco=(0,0)$, $\text{PS}_\bQ$ for $\qco=(0,\epsilon)$ with $\epsilon/2\pi \lesssim 0.125$, AFM+iCO for  $\qco=(0,\Qco)$, FIM+iCO for  $\qco=(\pi,\pi-\Qco)$, FIM for  $\qco=(\pi,\pi)$.
		\label{SM:fig:Uc}
}\end{figure}

\autoref{SM:Fig:chi} shows the momentum-dependent charge susceptibility around the AFM mean-field ground state for $n=1.1$ and different interactions $U,V$ on the high symmetry path of the BZ. The left panel shows different on-site interactions $U$ for a fixed interaction ratio of $V/U=0.1$. With increasing $U$, a divergence arises at $\QCO$ for $U_c\approx 11$, which implicates a charge instability of the MF ground state with emerging long-range order. For $U<U_c$, the charge susceptibility is peaked at the same wave vector, i.e., we detect short-range correlations with finite correlation length $\xi$ and the associated ordering vector $\QCO$ does not depend on $U$ as claimed in the main text. Thereby, $\xi$ grows monotonously with increasing interaction $U$, until it diverges at $U_c$, i.e., the onset of long-range order. The right panel of \autoref{SM:Fig:chi} displays $\chi_c$ for different interaction ratios $V/U$, where the on-site repulsion is tuned to the critical value $U\approx U_c$ for each ratio respectively.  This procedure allows to classify the emerging type of charge order (CO) by means of the ordering vector $\qco$, where the divergence occurs.  Notably, $\QCO$ is independent of $V$ as discussed in the main text. The interaction ratio $V/U$ has a decisive impact on which type of CO becomes dominant as exemplarily shown in \autoref{SM:Fig:chi}. 
The critical interaction $U_c$ as function of $V/U$ and associated types of emerging long-range order are displayed in \autoref{SM:fig:Uc} for four different fillings.
A comprehensive analysis in that fashion results in Figure 2 in the main text. As we claimed in that context, $U_c$ is enhanced for intermediate interaction ratios due to competing effects. Moreover, the overall critical interaction scale increases with the doping.

\section{Dynamic susceptibilities}\label{section:dynamicsus}
\begin{figure*}[h!]
	\begin{minipage}[b!]{0.48\linewidth}
		\flushleft
		\includegraphics[width=1.\linewidth]{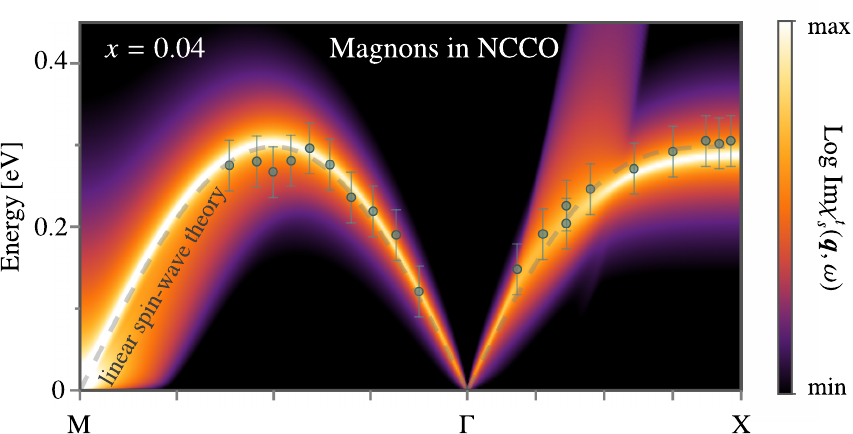}
		\caption{
			Magnon dispersion on the high-symmetry-path, which appears as enhanced line of $\chi_s^t$ in our theory ($U=8.2, V/U=0.1$), along with RIXS data in NCCO adopted from Ref.~\cite{lee2014asymmetry} displayed by green circles with error bars. 
			The absolute energy scale in our model is fitted to $t=0.33$~eV.
			The dashed line displays linear spin-wave theory with the velocity $c_s=0.83~\mathrm{eV}\AA$ adopted from Ref.~\cite{lee2014asymmetry}.}\label{SM:fig:magnon}
	\end{minipage}\hfill%
	\begin{minipage}[b!]{0.48\linewidth}
		\includegraphics[width=1.\linewidth]{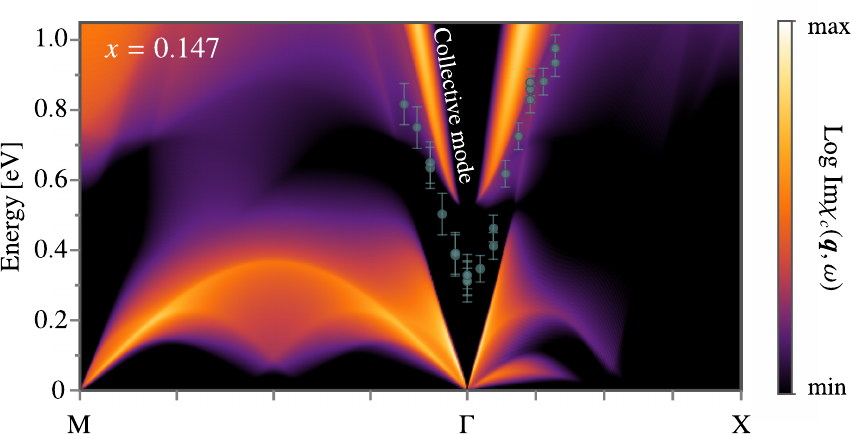}
		\caption{Gapped collective mode around the $\Gamma$-point in the charge susceptibility $\chi_c$ ($U=5.5, V/U=0.1$), along with RIXS data in NCCO adopted from Ref.~\cite{lee2014asymmetry} displayed by green circles with error bars. The absolute energy scale in our model is fitted to $t=0.52$~eV.\newline\newline}\label{fig:collectivemode}
	\end{minipage}%
\end{figure*}
We access finite frequencies via analytic continuation of the bosonic Matsubara frequency $\i\omega_n \rightarrow \omega+\i\eta$ with a convergence factor $\eta$. 
\autoref{SM:fig:magnon} shows the magnon dispersion that is encoded in the transversal spin susceptibility $\chi_s^t$ (fluctuating spin perpendicular to the MF spin) for $x=0.04$ as function of the momentum on the high-symmetry-path, i.e., $\Gamma=(0,0),\mathrm{X}=(0,\pi),\mathrm{M}=(\pi,\pi)$, along with RIXS data in NCCO \cite{lee2014asymmetry}. In contrast to the paramagnons shown in Fig. 4 in the main text, the dispersion is sharp and fits well to linear spin-wave theory of the AFM Heisenberg model, i.e.,
\begin{equation}
\omega= \sqrt{2}c_s\sqrt{1-\left(\frac{\cos q_x + \cos q_y}{2}\right)^2}\comma
\end{equation}
which is plotted by the dashed line. The magnetic modes are gapless at $\QSO=(\pi,\pi)$ and represent the Goldstone modes associated with the symmetry-broken magnetic mean-field ground states \cite{Seufert_2021}. 
\autoref{fig:collectivemode} displays the charge susceptibility $\chi_c$ for the same parameters as Fig. 4 in the main text that discussed paramagnons in the transversal spin susceptibility. We detect a collective mode around the $\Gamma$-point in the charge channel, where the gap size does not quantitatively match the gap of the experimentally observed charge mode \cite{lee2014asymmetry}, which could be due to its three-dimensional nature \cite{hepting2018three} while our model is 2D. Notably, the gap size decreases for lower interactions $U$ but too small values of $U$ lead to a mismatch for the paramagnon dispersion since the MF ground state becomes non-magnetic at $U\lesssim 3.2$ for $x=0.147$. 
\begin{figure}[h!]
	\includegraphics[width=0.999\linewidth]{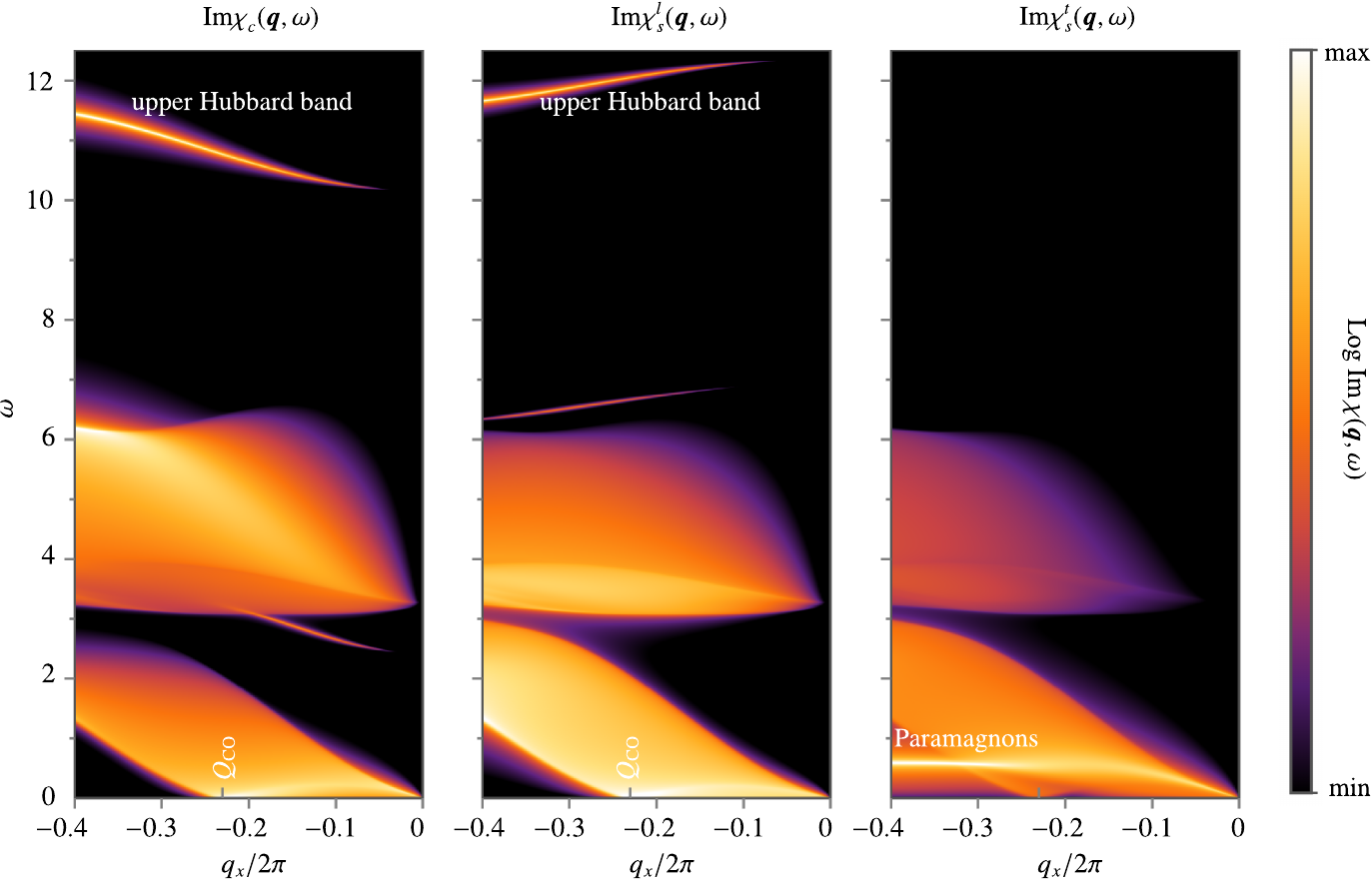}
	\caption{Dynamic charge susceptibility ($\chi_c$), longitudinal spin susceptibility ($\chi_s^l$) and transversal spin susceptibility ($\chi_s^t$) for $n=1.1$, $U=11$, $V/U=0.1$, $T=0.005$, $\eta=0.01$ and $\bq=(q_x,0)$.}\label{SM:fig:Xw}
\end{figure}
Finally, \autoref{SM:fig:Xw} shows dynamic charge and spin susceptibilities for $n=1.1$ in a wide range of frequencies that also captures the upper Hubbard band at $\omega\approx U$. $\chi_c$ and $\chi_s^l$ are dynamically coupled due to the AFM mean-field background, whereas $\chi^t_s$ is decoupled from the charge sector in the fluctuation matrix. In contrast to $\chi_s^t$, the sectors $\chi_c^l$ and $\chi_s^l$ feature significant spectral weight on the elastic line, i.e., $\omega=0$.

\end{document}